\documentclass[sigconf, nonacm, pdfa]{acmart}
\usepackage[a-2b]{pdfx}
\usepackage{xspace}
\usepackage{amsmath}
\usepackage{dsfont}
\usepackage{mathtools}
\usepackage[noend]{algpseudocode}
\usepackage{subcaption}
\usepackage{multirow}
\usepackage{pifont}
\usepackage[export]{adjustbox} 
\usepackage{enumitem}
\usepackage{balance}
\setlist{leftmargin=5.5mm}

\usepackage{fancyhdr}
\usepackage[absolute]{textpos}

\usepackage[symbol]{footmisc}
\usepackage[ruled, lined, linesnumbered, commentsnumbered, longend]{algorithm2e}
\SetAlFnt{\small}
\SetAlCapFnt{\normalsize}
\SetAlCapNameFnt{\normalsize}

\theoremstyle{definition}
\newtheorem{definition}{Definition}

\newcommand\vldbdoi{XX.XX/XXX.XX}
\newcommand\vldbpages{XXX-XXX}
\newcommand\vldbvolume{17}
\newcommand\vldbissue{3}
\newcommand\vldbyear{2023}
\newcommand\vldbauthors{\authors}
\newcommand\vldbtitle{\shorttitle} 
\newcommand\vldbavailabilityurl{URL_TO_YOUR_ARTIFACTS}
\newcommand\vldbpagestyle{plain}

\newcommand{\G}{$\boldsymbol{G}$\xspace}

\newcommand{\Lap}{$\boldsymbol{L}$\xspace}
\newcommand{\random}{\textit{Random}\xspace}
\newcommand{\kneighbor}{\textit{K-Neighbor}\xspace}
\newcommand{\rankdegree}{\textit{Rank Degree}\xspace}
\newcommand{\localdegree}{\textit{Local Degree}\xspace}
\newcommand{\spanning}{\textit{Spanning Forest}\xspace}

\newcommand{\tspanner}{\textit{t-Spanner}\xspace}
\newcommand{\tspanners}{\textit{t-Spanners}\xspace}

\newcommand{\forestfire}{\textit{Forest Fire}\xspace}
\newcommand{\lspar}{\textit{L-Spar}\xspace}
\newcommand{\gspar}{\textit{G-Spar}\xspace}
\newcommand{\localsimilarity}{\textit{Local Similarity}\xspace}
\newcommand{\scan}{\textit{SCAN}\xspace}
\newcommand{\er}{\textit{ER}\xspace}
\newcommand{\erw}{\textit{ER-weighted}\xspace}
\newcommand{\eruw}{\textit{ER-unweighted}\xspace}

\newcommand{\facebook}{{\fontfamily{lmss}\selectfont ego-Facebook}\xspace}
\newcommand{\twitter}{{\fontfamily{lmss}\selectfont ego-Twitter}\xspace}
\newcommand{\human}{{\fontfamily{lmss}\selectfont human\_gene2}\xspace}
\newcommand{\dblp}{{\fontfamily{lmss}\selectfont com-DBLP}\xspace}
\newcommand{\amazon}{{\fontfamily{lmss}\selectfont com-Amazon}\xspace}
\newcommand{\enron}{{\fontfamily{lmss}\selectfont email-Enron}\xspace}

\newcommand{\astroph}{{\fontfamily{lmss}\selectfont ca-AstroPh}\xspace}
\newcommand{\hepph}{{\fontfamily{lmss}\selectfont ca-HepPh}\xspace}
\newcommand{\berkstan}{{\fontfamily{lmss}\selectfont web-BerkStan}\xspace}
\newcommand{\google}{{\fontfamily{lmss}\selectfont web-Google}\xspace}
\newcommand{\notredame}{{\fontfamily{lmss}\selectfont web-NotreDame}\xspace}
\newcommand{\stanford}{{\fontfamily{lmss}\selectfont web-Stanford}\xspace}
\newcommand{\reddit}{{\fontfamily{lmss}\selectfont Reddit}\xspace}
\newcommand{\proteins}{{\fontfamily{lmss}\selectfont ogbn-proteins}\xspace}

\newcommand{\cmark}{\text{\ding{51}}}
\newcommand{\xmark}{\text{\ding{55}}}
\newcommand{\omark}{\text{\ding{108}}}
\definecolor{ForestGreen}{RGB}{34,139,34}
\definecolor{myyellow}{RGB}{181, 181, 27}
\newcommand{\greencheck}{{\color{ForestGreen}\cmark}}
\newcommand{\yellowcheck}{{\color{myyellow}\cmark}}
\newcommand{\redcheck}{{\color{red}\xmark}}
\newcommand{\yellowo}{{\color{myyellow}\omark}}

\begin{document}
\title{Demystifying Graph Sparsification Algorithms in Graph Properties Preservation}

\begin{textblock}{16}(3,1)
{\normalsize \normalfont \textit{Authors' version; to appear in the Proceedings of 50th International Conference on Very Large Databases (VLDB 2024).}}
\end{textblock}

\author{Yuhan Chen}
\affiliation{%
  \institution{University of Michigan}
}
\email{chenyh@umich.edu}

\author{Haojie Ye}
\affiliation{%
  \institution{University of Michigan}
}
\email{yehaojie@umich.edu}

\author{Sanketh Vedula}
\affiliation{%
  \institution{Technion}
}
\email{sanketh@cs.technion.ac.il}

\author{Alex Bronstein}
\affiliation{%
  \institution{Technion}
}
\email{bron@cs.technion.ac.il}

\author{Ronald Dreslinski}
\affiliation{%
  \institution{University of Michigan}
}
\email{rdreslin@umich.edu}

\author{Trevor Mudge}
\affiliation{%
  \institution{University of Michigan}
}
\email{tnm@umich.edu}

\author{Nishil Talati}
\affiliation{%
  \institution{University of Michigan}
}
\email{talatin@umich.edu}


\begin{abstract}
Graph sparsification is a technique that approximates a given graph by a sparse graph with a subset of vertices and/or edges. The goal of an effective sparsification algorithm is to maintain specific graph properties relevant to the downstream task while minimizing the graph's size. Graph algorithms often suffer from long execution time due to the irregularity and the large real-world graph size. Graph sparsification can be applied to greatly reduce the run time of graph algorithms by substituting the full graph with a much smaller sparsified graph, without significantly degrading the output quality. However, the interaction between numerous sparsifiers and graph properties is not widely explored, and the potential of graph sparsification is not fully understood.

In this work, we cover 16 widely-used graph metrics, 12 representative graph sparsification algorithms, and 14 real-world input graphs spanning various categories, exhibiting diverse characteristics, sizes, and densities. We developed a framework to extensively assess the performance of these sparsification algorithms against graph metrics, and provide insights to the results. Our study shows that there is no one sparsifier that performs the best in preserving all graph properties, e.g. sparsifiers that preserve distance-related graph properties (eccentricity) struggle to perform well on Graph Neural Networks (GNN). This paper presents a comprehensive experimental study evaluating the performance of sparsification algorithms in preserving essential graph metrics. The insights inform future research in incorporating matching graph sparsification to graph algorithms to maximize benefits while minimizing quality degradation. Furthermore, we provide a framework to facilitate the future evaluation of evolving sparsification algorithms, graph metrics, and ever-growing graph data.
\end{abstract}
\maketitle

\pagestyle{\vldbpagestyle}
\begingroup\small\noindent\raggedright\textbf{PVLDB Reference Format:}\\
\vldbauthors. \vldbtitle. PVLDB, \vldbvolume(\vldbissue): \vldbpages, \vldbyear.\\
\href{https://doi.org/\vldbdoi}{doi:\vldbdoi}
\endgroup
\begingroup
\renewcommand\thefootnote{}\footnote{\noindent
This work is licensed under the Creative Commons BY-NC-ND 4.0 International License. Visit \url{https://creativecommons.org/licenses/by-nc-nd/4.0/} to view a copy of this license. For any use beyond those covered by this license, obtain permission by emailing \href{mailto:info@vldb.org}{info@vldb.org}. Copyright is held by the owner/author(s). Publication rights licensed to the VLDB Endowment. \\
\raggedright Proceedings of the VLDB Endowment, Vol. \vldbvolume, No. \vldbissue\ %
ISSN 2150-8097. \\
\href{https://doi.org/\vldbdoi}{doi:\vldbdoi} \\
}\addtocounter{footnote}{-1}\endgroup

\ifdefempty{\vldbavailabilityurl}{}{
\vspace{.3cm}
\begingroup\small\noindent\raggedright\textbf{PVLDB Artifact Availability:}\\
The source code, data, and/or other artifacts have been made available at \url{https://github.com/yuhanchan/sparsification}.
\endgroup
}

\section{Introduction}
Graphs are ubiquitous because of their great expressiveness and flexibility. Graphs can be used to represent complex relationships between individuals (vertices in the graph) by making connections (edges in the graph). Graphs are widely used to represent data in various application domains, e.g. social networks~\cite{linton2004}, citation and communication networks~\cite{newman2004}, chemical and biological networks~\cite{Jeong2000}, etc. Many algorithms are also developed to exploit the abundant features that graphs provide, e.g. Dijkstra's algorithm~\cite{dijkstra1959}, Ford-Fulkerson algorithm~\cite{fulkerson1956}, Graph Neural Networks~\cite{kipf2016}, etc.

Despite their usefulness, graphs are often inefficient to work with due to memory irregularity. Many works are proposed to tackle the problem~\cite{song2018, talati2022,chen2023}, however, most works develop dedicated software or hardware solutions for a small set of graph algorithms, which leads to a high design cost and limited applicability. In this work, we investigate graph sparsification, a generally applicable technique to reduce the amount of work in graph algorithms.

Graph sparsification is a technique to approximate a given graph by a sparse graph that preserves certain properties of the graph. This way we can execute the downstream task on the sparsified graph to improve run time. An ideal sparsification algorithm needs to achieve a high prune rate while keeping the behavior of the downstream task as close to that of the original full graph.

There are many sparsification algorithms with different focuses on the graph properties to be preserved, and of different complexity. There are also many graph metrics that different graph-centric algorithms rely on. However, with a large number of sparsification algorithms and graph metrics, the connections between sparsifiers and their performance in preserving the graph metrics are missing. 

In this work, we extensively investigate 12 graph sparsification algorithms and evaluate their performance in preserving 16 widely-used graph metrics in mutiple groups. We also cover 14 real-world graphs spanning various categories, with diverse characteristics, sizes, and densities. Our findings reveal that no single sparsifier does the best in preserving all graph properties, and it is important to select appropriate sparsifiers based on the downstream task. 

In summary, we make the following contributions in this work:

\begin{itemize}
    \item We summarize the most widely-used graph metrics and the most representative graph sparsification algorithms, and dig into the algorithmic details for a better understanding.
    \item We build a framework to perform graph sparsification, and evaluate their performance on various graph metrics at different prune rates. The framework is open-source and extendable to future sparsification algorithms, graph metrics, and graphs.
    \item We perform N-to-N evaluation on the sparsification algorithms and graph metric, give a comprehensive breakdown of the performance, and provide insights with the results.
\end{itemize}

\section{overview}
\subsection{Preliminaries}
In this section, we introduce the basic notions used in this paper.

Consider a graph $\boldsymbol{G} = (\mathcal{V}, \mathcal{E}, \boldsymbol{w})$, where $\mathcal{V}$ and $\mathcal{E}$ denotes the set of vertices and edges in $\boldsymbol{G}$ respectively, and $\boldsymbol{w}$ denotes the weights of the edges. A graph can be either directed or undirected. In a directed graph, each edge has a source and a destination vertex, while an undirected graph implies a bidirectional relationship. Furthermore, a graph can be weighted or unweighted; in an unweighted graph, all edges have a default weight of 1. $|\mathcal{V}|$, $|\mathcal{E}|$ represent the number of vertices and edges, respectively. A graph is considered \textit{connected} if a path exists between any pair of vertices~\cite{wolframConnected}. We denote the adjacency matrix by $\boldsymbol{A}$, with the entries in $\boldsymbol{A}$ defined as:
\begin{equation*} \label{adjMat}
    \boldsymbol{A}_{ij}=
    \begin{cases}
      \boldsymbol{w}_{i\rightarrow j} & \text{if}\ e_{ij}\in\mathcal{E}, \\
      0 & \text{otherwise}.
    \end{cases}
\end{equation*}
We denote the graph Laplacian matrix by \Lap defined as follows:
\begin{equation*} \label{lapMat}
    \boldsymbol{L}_{ij} =\boldsymbol{D}-\boldsymbol{A}=
    \begin{cases}
      deg({v}_{i}) & \text{if}\ i=j, \\
      -\boldsymbol{w}_{i \to j} & \text{if}\ {e}_{ij} \in \mathcal{E}, \\
      0 & \text{otherwise}.
    \end{cases}
\end{equation*}
Note that we only consider Laplacian matrices for undirected graphs, thus the graph Laplacian is a positive semi-definite matrix.
We now present a formal definition of the graph sparsification problem.
\begin{definition}[Graph Sparsification]
Let $\boldsymbol{G} = (\mathcal{V}, \mathcal{E}, \boldsymbol{w})$ be a given graph. A \textit{sparsified subgraph} $\boldsymbol{H} = (\mathcal{V}, \mathcal{\Tilde{E}}, \boldsymbol{\Tilde{w}})$ is constructed such that $|\mathcal{\Tilde{E}}| = (1-\rho) |\mathcal{E}|$. The function $f$ that creates $\boldsymbol{H}$ from $\boldsymbol{G}$, $\boldsymbol{H} = f(\boldsymbol{G})$, is called a \textit{graph sparsification algorithm} (also referred to as a \textit{sparsifier}), while $\rho$ is defined as the \textit{prune rate}.
\end{definition}
In this study, our focus is solely on edge sparsification, implying that we maintain the original vertex set while selecting a subset of edges. This approach is adopted for several reasons: 1) the edge set typically possesses a significantly larger size than the vertex set and contains more redundant information, 2) the majority of sparsification algorithms focus on pruning edges rather than vertices, and 3) most graph metrics require the complete set of vertices for evaluating the performance of sparsification algorithms.

\subsection{Graph Metrics} \label{sec:metrics}
\subsubsection{Basic Metrics} \hfill\\
This section introduces some fundamental graph metrics.

\textbf{Degree Distribution.} The degree of a vertex is defined as the number of edges incident to it. The degree distribution provides a comprehensive perspective on the graph's structure, enabling the classification of different types of graphs. For instance, a randomly generated graph might exhibit a uniform degree distribution, whereas a real-world social network has a power-law distribution.

\textbf{Laplacian Quadratic Form.} 
This is defined as $\boldsymbol{x}^T \boldsymbol{L} \boldsymbol{x}$, where $\boldsymbol{L}$ represents the graph Laplacian, and $\boldsymbol{x} \in \mathbb{R}^{|\mathcal{V}|}$ is an arbitrary vector. The Laplacian quadratic form is a fundamental quantity in graph theory~\cite{Bondy1976}, and it facilitates the analysis of various graph properties, including connectivity and spectral characteristics~\cite{mikhail2001}.

\subsubsection{Distance Metrics} \hfill\\
\label{sec:distanceMetrics}
This section includes a collection of metrics associated with the pairwise distances between vertices in graphs.

\textbf{All Pairs Shortest Path (APSP).} APSP measures the minimum distance between any pair of source vertex $u$ and destination vertex $v$. Breadth-First Search (BFS) and Dijkstra's algorithm~\cite{dijkstra1959} are often used to determine APSP. Distance captures the proximity between two vertices. APSP are used in various domains such as data center network design~\cite{Curtis2011} and urban service system planning~\cite{larson1981}.

\textbf{Diameter.} The diameter of a graph \G is defined as the maximum distance between any pair of vertices $u$ and $v$. If \G is disconnected, its diameter is considered infinite. The diameter is useful in various applications, including transportation network planning~\cite{geoff2018} and the analysis of routing and communication network quality~\cite{richard2004}.

\textbf{Vertex Eccentricity.} 
Vertex eccentricity is defined as the length of the longest shortest path from a source vertex $s$ to all other vertices in \G. Note that the minimum eccentricity is the graph radius, and the maximum eccentricity is the graph diameter. Vertex eccentricity is infinite for disconnected graphs. It identifies vertices located near the geometrical center of the graph.
Vertex eccentricity has practical applications in identifying network periphery in routing network~\cite{magoni2001, takes2013}. Or identifying proteins readily functionally reachable by other components in protein networks.~\cite{takes2013, GA2011}.

\subsubsection{Centrality Metrics} \label{sec:metric_centrality} \hfill\\
Centrality measures are a set of metrics employed to assess the significance or ranking of vertices in various manners.

\textbf{Betweenness.} 
Betweenness centrality for vertex $v$ is defined as
\begin{equation*} \label{BC}
    C_\text{betweenness}(v) = \sum_{s\neq v\neq t}\frac{\sigma_{st}(v)}{\sigma_{st}}.
\end{equation*}
Here, $\sigma_{st}$ denotes the total number of shortest paths from vertex $s$ to $t$, while $\sigma_{st}(v)$ refers to the number of shortest paths passing through $v$. The underlying intuition suggests that vertices appearing on numerous shortest paths exhibit high betweenness centrality. It can be employed to identify hubs in a transportation network~\cite{jan2006} or to identify important vertices (people) in social networks~\cite{burt1992}.

\textbf{Closeness.} 
Closeness centrality~\cite{bergamini2017} of a vertex $v$ is defined as 
\begin{equation*} \label{CC}
    C_\text{closeness}(v) = \frac{1}{\sum_u d(u,v)}.
\end{equation*}
Here, $d(u,v)$ represents the shortest distance between vertices $u$ and $v$. The underlying intuition is that vertices with a shorter average distance to all other reachable vertices exhibit high closeness centrality. It can identify essential genes in protein-interaction networks~\cite{hahn2004} or crucial metabolites in metabolic networks~\cite{ma2003}.

\textbf{Eigenvector.} 
Eigenvector centrality of a vertex $v$ is defined as 
\begin{equation*} \label{EC}
    C_\text{eigenvector}(v) = \frac{1}{\lambda}\sum_{u\in N(v)}C_{eigenvector}(u).
\end{equation*}
where $N(v)$ is the neighbour of $v$, and $\lambda$ is the greatest eigenvalue of the adjacency matrix $\boldsymbol{A}$.
Eigenvector centrality measures the influence of a vertex~\cite{WikiEigenCent}. A high eigenvector score means that a vertex is connected to many vertices whose eigenvector scores are also high~\cite{Newman2016}.
Google's PageRank~\cite{Page1999} and Katz centrality are two variants of eigenvector centrality. We discuss Katz centrality in the next paragraph and PageRank in Section~\ref{sec:highlevel}. Eigenvector centrality is useful for assessing opinion influence in sociology and economics~\cite{rapanos2023}, or the firing rate of neurons in neuroscience~\cite{fletcher2018}.

\textbf{Katz.} 
Katz centrality quantifies the influence of a vertex by considering the number of immediate neighbors and vertices connected to those immediate neighbors~\cite{Katz1953}. Distant neighbors are penalized by an attenuation factor $\alpha^k$, where $k$ represents the hop distance from the central vertex. 
In this paper, we use $\alpha=1/(max(degree)+1)$.
The eigenvector centrality is defined as
\begin{equation*} \label{KatzC}
    C_{Katz}(v) = \sum_{k}\sum_{u}\alpha^k(\boldsymbol{A}^k)_{uv}.
\end{equation*}

\subsubsection{Clustering Metrics} \label{sec:cluster_metrics}\hfill\\
Graph clustering groups vertices into communities, ensuring dense connections within communities and sparse connections between communities. This section covers graph clustering-related metrics.

\textbf{Number of communities.} 
The most basic metric in graph clustering is the number of communities. For graphs with a known number of communities $k$, certain clustering algorithms, such as k-means~\cite{lloyd1982}, can construct exactly $k$ communities. Alternatively, some algorithms like agglomerative clustering~\cite{daniel2011} and DBSCAN~\cite{martin1996} can automatically determine the optimal number of clusters.

\textbf{Local Clustering Coefficient (LCC).} 
LCC of a vertex $v$ represents the proportion of pairs of neighbors of $v$ that are connected. It evaluates the density of connections among the neighbors of a vertex~\cite{wikiClustering2023}. The LCC is defined as follows:
\begin{equation*} \label{eq:LCC}
\small
    LCC(v) = \frac{|e_{jk}:j,k\in N_v, e_{jk}\in E|}{\alpha k_v(k_v-1)}.
\end{equation*}
where $N_v$ denotes the set of neighbors of the vertex $v$, and $k_v$ is the number of neighbors of vertex $v$. Here, $\alpha=1$ for directed graphs, and $\alpha=0.5$ for undirected graphs. LCC, originally proposed by Watts and Strogatz, is used to determine whether a graph is a small-world network~\cite{watts1998}. \textbf{Mean clustering coefficient (MCC)} is the mean of the local clustering coefficient of all vertices.

\textbf{Global Clustering Coefficient (GCC).} GCC~\cite{Luce1949AMO} measures the fraction of closed triplets in all triplets. A triplet of nodes can consist of two (open) or three (closed) undirected edges~\cite{wikiClustering2023}.
\begin{equation*} \label{eq:GCC}
\small
    GCC(v) = \frac{\# Closed\ triplets}{\#All\ triplets}.
\end{equation*}

\textbf{Clustering F1 score.} 
The F1 score can be employed to assess the similarity between a given clustering and a reference clustering~\cite{mallawaarachchi_2020}. Suppose we have $k$ clusters $C_i$ ($i \in [1, k]$) obtained from a specific algorithm for graph \textit{G} and $s$ reference clusters $R_j$ ($j \in [1, s]$) that we aim to compare with. Note that $s$ may not be equal to $k$.
The following matrix illustrates the relationship between $C_i$ and $R_j$:
\begin{equation*} \label{CRmatrix}
\small
\begin{matrix}
                 & \boldsymbol{R_1} & \boldsymbol{R_2} & \boldsymbol{...} & \boldsymbol{R_s} \\
\boldsymbol{C_1} & a_{11} & a_{12} & ... & a_{1s} \\
\boldsymbol{C_2} & a_{21} & a_{22} & ... & a_{2s} \\
\boldsymbol{...} \\
\boldsymbol{C_k} & a_{k1} & a_{k2} & ... & a_{ks} 
\end{matrix}
\end{equation*}
In this matrix, $a_{ij}$ represents the number of vertices shared between cluster $C_i$ and reference cluster $R_j$.
The precision and recall of the clustering are defined as follows:
\begin{equation*} \label{eq:clusterPrecisionRecall}
\small
    Precision = \frac{\sum_{i\in[1,k]}max_j\{a_{ij}\}}{\sum_{i\in[1,k]}\sum_{j\in[1,s]}a_{ij}}, \text{\ \ } Recall = \frac{\sum_{i\in[1,k]}max_j\{a_{ij}\}}{n}
\end{equation*}
Subsequently, the F1 score for clustering is defined as:
\begin{equation*} \label{clusterF1}
    F1 = 2\times\frac{Precision \times Recall}{Precision + Recall}
\end{equation*}
The F1 score ranges from 0 to 1, where a higher value indicates greater similarity between the clustering $\boldsymbol{C}$ and the reference $\boldsymbol{R}$.

\begin{table}[h!]
\centering
\scriptsize
\begin{tabular}{l|ccc}
 \textbf{Metric} & \textbf{Directed} & \textbf{Weighted} & \textbf{Unconnected}\\ \hline \hline
\textbf{Degree Dist.} & \greencheck & \yellowo$^\dag$ & \greencheck \\
\textbf{Diameter} & \greencheck & \greencheck & \yellowcheck$^\ddag$\\
\textbf{Eccentricity} & \greencheck & \greencheck & \yellowcheck$^\ddag$\\
\textbf{APSP} & \greencheck & \greencheck & \yellowcheck$^\ddag$ \\
\textbf{Betweenness Cent.} & \greencheck & \greencheck & \greencheck\\
\textbf{Closeness Cent.} & \greencheck & \greencheck & \greencheck\\
\textbf{Eigenvector Cent.} & \yellowcheck$^{*}$ & \greencheck & \greencheck\\
\textbf{Katz Cent.} & \greencheck & \greencheck & \greencheck\\
\textbf{\#Communities} & \redcheck & \greencheck & \greencheck\\
\textbf{LCC} & \greencheck & \yellowo$^\dag$ & \greencheck\\
\textbf{MCC} & \greencheck & \yellowo$^\dag$ & \greencheck\\
\textbf{GCC} & \greencheck & \yellowo$^\dag$ & \greencheck\\
\textbf{Clustering F1 Sim} & \redcheck & \greencheck & \greencheck\\
\textbf{PageRank} & \greencheck & \greencheck & \greencheck\\
\textbf{Min-cut/Max-flow} & \greencheck & \greencheck & \yellowcheck$^\ddag$\\
\textbf{GNN} & \greencheck & \greencheck & \greencheck \\ \hline \hline
\multicolumn{4}{l}{
  \begin{minipage}{8cm}
    \footnotesize $*$ For directed graphs, the left eigenvector is used. A left eigenvector is an eigenvector satisfies $X_L \boldsymbol{A} = \lambda_L X_L$, where a right (by default) eigenvector satisfies $\boldsymbol{A} X_R = \lambda_R X_R$ \\
    \footnotesize $\dag$ Weight not used, same as unweighted. \\
    \footnotesize $\ddag$ In unconnected graphs, pair-wise distance can be infinite, and min-cut max-flow can be zero if two terminals selected are in different communities. We exclude these pairs in the evaluation.\\
  \end{minipage}
}\\
\end{tabular}
\caption{Metrics' applicability to types of graphs.}
\label{tab:metric}
\end{table}

\subsubsection{Application-level Metrics} \label{sec:highlevel} \hfill\\
In this section, we discuss metrics that are used in applications. 

\textbf{PageRank.} 
PageRank, initially designed to rank web pages~\cite{Page1999}, serves as a foundational algorithm for Google's search engine. The underlying concept suggests that pages linked by numerous important pages bear greater significance. PageRank computation typically employs the power method. Each page (vertex) is assigned an initial score and iteratively calculates a new score by adding up $1/k$ of the scores of pages linked to it, where $k$ represents the number of outgoing links from the source page. Eventually, the computation converges, and the score of each page indicates its importance within the network. The primary distinction between PageRank and eigenvector centrality (\S~\ref{sec:metric_centrality}) lies in PageRank's specificity for web-page ranking, incorporating $1/k$ factor and additional parameters like damping factor~\cite{brin1998} for better robustness and accuracy, while eigenvector centrality is more suitable for general graph analysis, not necessarily involve directed or weighted graphs.

\textbf{Min-cut and Max-flow.}
In graph theory, a cut refers to the partitioning of a graph's vertices into two disjoint subsets~\cite{wikicut}. A minimum $s$-$t$ cut, or min-cut, represents the cut with the smallest total weight of edges that disconnect the source vertex ${s}$ from the sink vertex ${t}$. The maximum flow, or max-flow, denotes the maximum amount of flow that can traverse from the source vertex ${s}$ to the sink vertex ${t}$, where the edge weight represents the flow capacity. The max-flow and min-cut problems are equivalent, as the maximum flow a network can accommodate is constrained by the network's narrowest intersection, which is the min-cut. Min-cut and max-flow can be applied to identify bottlenecks in water networks, road networks, or electrical networks~\cite{ramesh2017, ahuja1993}.

\textbf{Graph Neural Networks (GNNs).}
GNNs~\cite{scarselli2009} are neural networks that operate on graphs. GNNs learn from the graph structure by aggregating information from neighboring vertices or edges and feeding the information to multi-layer perception (MLP) layers for training. Some famous GNN models include Graph Convolutional Network (GCN), Graph Attention Network (GAT), and ChebNet~\cite{kipf2016,defferrard2016,petar2017}. GNN can be used for classification or prediction on vertex level, edge level, or graph level tasks~\cite{zhou2019,li2019,demir2021}. 

We summarize the graph metrics discussed and their applicability to different types of graphs in table~\ref{tab:metric}.
\subsection{Graph Sparsification Algorithms}

\begin{table*}[ht!]
\scriptsize
\begin{tabular}{l|ccc|ccc|l}
 \textbf{Sparsifier} & \textbf{\small Directed?} & \textbf{\small Weighted?} & \textbf{\small Unconnected?} & \textbf{\small PRC$^\S$} & \textbf{\small Weight Change} & \textbf{\small Deterministic?} & \textbf{\small Complexity$^{**}$} \\ \hline \hline
\textbf{\random (RN)} & \greencheck & \greencheck & \greencheck & \greencheck & \redcheck & \redcheck & $\mathcal{O}(\rho |\mathcal{E}|)$ \\
\textbf{\kneighbor (KN)} & \yellowcheck$^*$ & \greencheck & \greencheck & \yellowcheck$^\ddag$ & \redcheck & \redcheck & $\mathcal{O}(|\mathcal{E}|)$ \\
\textbf{\rankdegree (RD)} & \yellowcheck$^*$ & \greencheck & \greencheck & \yellowcheck$^\ddag$ & \redcheck & \redcheck & $\mathcal{O}(\rho |\mathcal{E}|) - \mathcal{O}(\rho |\mathcal{E}|)log(\rho |\mathcal{E}|)$ \\
\textbf{\localdegree (LD)} & \yellowcheck$^*$ & \greencheck & \greencheck & \yellowcheck$^\ddag$ & \redcheck & \greencheck & $\mathcal{O}(|\mathcal{E}|) - \mathcal{O}(|\mathcal{E}|log(|\mathcal{E}|))$ \\
\textbf{\spanning (SF)} & \redcheck & \greencheck & \greencheck & \redcheck & \redcheck & \greencheck & $\mathcal{O}(|\mathcal{E}| log(|\mathcal{V}|))$ \\
\textbf{\tspanner (SP-t)} & \redcheck & \greencheck & \greencheck & \redcheck & \redcheck & \greencheck & $\mathcal{O}(|\mathcal{V}|^2log(|\mathcal{V}|))$ \\
\textbf{\forestfire (FF)} & \greencheck & \greencheck & \yellowcheck$^\dag$ & \yellowcheck$^\ddag$ & \redcheck & \redcheck & $\mathcal{O}(r|\mathcal{E}|)$ \\
\textbf{\lspar (LS)} & \yellowcheck$^*$ & \greencheck & \greencheck & \yellowcheck$^\ddag$ & \redcheck & \greencheck & $\mathcal{O}(k|\mathcal{E}|)$ \\
\textbf{\gspar (GS)} & \yellowcheck$^*$ & \greencheck & \greencheck & \greencheck & \redcheck & \greencheck & $\mathcal{O}(k|\mathcal{E}|)$ \\
\textbf{\localsimilarity (LSim)} & \yellowcheck$^*$ & \greencheck & \greencheck & \yellowcheck$^\ddag$ & \redcheck & \greencheck & $\mathcal{O}(|\mathcal{E}|)$ \\
\textbf{\scan} & \yellowcheck$^*$ & \greencheck & \greencheck & \greencheck & \redcheck & \greencheck & $\mathcal{O}(|\mathcal{E}|)$ \\
\textbf{\er} & \redcheck & \greencheck & \greencheck & \greencheck & \greencheck & \redcheck & $\mathcal{O}(|\mathcal{E}| log(|\mathcal{V}|)^3)$ \\ \hline \hline
\multicolumn{8}{l}{
  \begin{minipage}{16cm}
    \footnotesize $\S$ Prune Rate Control. Whether the sparsifier has fine-grain, coarse-grain, or no control over the prune rate. \\ 
    \footnotesize $*$ Need to specify using in-degree or out-degree, in this work, we use out-degree.\\
    \footnotesize $**$ $|\mathcal{V}|=$\#Vertices, $|\mathcal{E}|=$\#Edges, $\rho=$prune rate, $r=$burnt ratio, $k=$\#minwise hash. It can be a range for some sparsifiers due to different optimal algorithms can be used according to graph properties.\\
    \footnotesize $\dag$ Seeds are randomly selected, thus edges from communities with fewer vertices are less likely to be included.\\
    \footnotesize $\ddag$ Subject to constraint. Indirect or coarser grain control, or has an upper limit for prune rate. \\
\end{minipage}
}\\
\end{tabular}
\caption{Sparsifiers' applicability to types of graphs, and characteristics. Note that all sparsifiers work for undirected, unweighted, and connected graphs because they are special cases of directed, weighted, and unconnected graphs, so are not listed. Deterministic means whether the sparsifier generates the same sub-graph every time.}
\label{tab:sparsifiers}

\end{table*}

In this section, we discuss graph sparsification algorithms evaluated in this work; they constitute the most widely used and representative sparsification algorithms.

\subsubsection{\random Sparsifier}\hfill\\
Arguably the simplest way to sparsify the graph is by randomly sampling a subset of edges to keep in the sparsified graph. We refer to this as the \random sparsifier. It samples all edges in the graph with equal probability and thus can be used to preserve vertex-relative (distribution-based and ranking-based) properties. \random sparsifier is employed in GraphSAGE for neighbor sampling~\cite{hamilton2017}. 

\subsubsection{\kneighbor Sparsifier}\hfill\\
\kneighbor sparsifier~\cite{sadhanala2016} selects $k$ edges for each vertex, and if a vertex has less than $k$ vertices, all of its edges are included. The edges are selected with probability proportional to their weights (uniform for unweighted graphs). It can be used in Laplacian smoothing~\cite{sadhanala2016}. \kneighbor guarantees each vertex has at least $k$ edges, so it can be applied if the downstream task requires high graph connectivity.

\subsubsection{\rankdegree Sparsifier}\hfill\\
\rankdegree sparsifier~\cite{elli2016} starts with selecting a random set of ``seed'' vertices. Subsequently, the vertices with edges to the seed vertices are ranked according to their degree in descending order. The edges connecting each seed vertex to its top-ranked neighbors are selected and incorporated into the sparsified graph. The recently added nodes in the graph serve as new seeds to search for additional edges. This process continues until the target sparsification limit is reached. \rankdegree biases to high-degree vertices, which are considered the hub vertices in a graph, so it excels at keeping edges incident to the important vertices in graphs.
\subsubsection{\localdegree Sparsifier}\hfill\\
Similar to the \rankdegree sparsifier, the \localdegree sparsifier~\cite{hamann2016} preserves edges incident to high-degree vertices, but in a deterministic manner. For each vertex, \localdegree incorporates edges to the top $deg(v)^\alpha$ neighbors ranked by their degree in descending order, where $\alpha \in [0, 1]$ controls the degree of sparsification. Another difference compared to \rankdegree is that \localdegree sparsifier makes sure each vertex will have at least 1 edge, so \localdegree sparsifier is a good choice when one desires to keep both graph connectivity and edges incident to important vertices.

\subsubsection{\spanning}\hfill\\
A spanning tree is a subgraph that constitutes a tree (a connected graph without a cycle~\cite{wikiTree}) and includes all the vertices in the graph~\cite{wikiSpanningTree}. A \spanning consists of multiple spanning trees. Kruskal's algorithm~\cite{Kruskal1956} and Prim's algorithm~\cite{Prim1957} can be used to construct a \spanning. Although it is not strictly a sparsifier, as the prune rate cannot be controlled, we include \spanning because it reduces the size of graphs and is a fundamental notion in graph theory. \spanning is helpful when one strictly wants to keep the graph connectivity the same as the original graph.

\subsubsection{\tspanner}\hfill\\
A spanner is a subgraph that approximates the pairwise distances between vertices in the original graph. A \tspanner is defined as a subgraph such that any pairwise distance is at most $t$ times the distance in the original graph, which can be formally expressed as:
\begin{equation*} \label{eq:tspanner}
\forall u,v\in \boldsymbol{V},d_{\boldsymbol{H}}(u,v)\le td_{\boldsymbol{G}}(u,v)
\end{equation*}
In this equation, ${t} (> 1)$ denotes the stretch factor. A greedy algorithm~\cite{ingo1990} is employed for constructing t-spanners. This algorithm starts with an empty edge set and then iteratively adds the edge $e_{uv}$ if the distance $d_H(u,v)$ between the vertices $u$ and $v$ in the current graph exceeds t times the weight of $e_{uv}$. The process continues until all edges have been considered. In addition to strictly keeping the graph connectivity, \tspanner also provides a better guarantee on the pair-wise distances between vertices and is a better choice than \spanning when such property is desired.

\subsubsection{\forestfire}\hfill\\
The \forestfire model is a generative model for graphs, originally proposed by Leskovec et al.~\cite{Leskovec2006}. The concept involves constructing the graph by adding one vertex at a time and forming edges to certain subsets of the existing vertices. When a new vertex $u$ is added to the graph, it connects to an existing vertex $v$ in the graph. Subsequently, it ``spreads'' from $v$ to other vertices in the graph with a certain predefined probability, creating edges between $v$ and the newly discovered vertices. This process assembles ``burning'' through edges probabilistically, hence the name \forestfire\cite{Leskovec2006}.

\subsubsection{Similarity-based sparsifiers}\hfill\\
Similarity-based sparsifiers constitute a group of sparsifiers based on similarities between vertices measured by specific metrics.

Jaccard similarity~\cite{murphy1996} measures the similarity between two sets by computing the portion of shared neighbors between two nodes ($u$ and $v$), as defined below:
\begin{equation*} \label{eq:jaccardSim}
    JaccardSimilarity({u}, {v}) = \frac{|\mathcal{N}({u})\bigcap\mathcal{N}({v})|}{|\mathcal{N}({u})\bigcup\mathcal{N}({v})|} 
\end{equation*}
The \textit{Jaccard score} of an edge is the Jaccard similarity between two constituent vertices of the edge. Once Jaccard scores are computed, they can be used to perform similarity-based sparsifications.

\textbf{Global Sparsifiers.}
Global sparsifiers select edges based on similarity scores globally. \textit{global Jaccard sparsifier} (\gspar) sorts the Jaccard scores globally and then selects the edges with the highest similarity score. \scan~\cite{xu2007} uses structural similarity measures to detect clusters, hubs, and outliers. 
The SCAN similarity score is a modified version of the Jaccard score, defined as follows:
\begin{equation*} \label{eq:SCANscore}
    SCANSimilarity({u}, {v}) = \frac{|\mathcal{N}({u})\bigcap\mathcal{N}({v})|+1}{\sqrt{(deg(u)+1)(deg(v)+1)}} 
\end{equation*}
Once the scores are computed, the edges in the sparsified graph are included from high-score edges to low-score edges.

\textbf{Local Sparsifiers.}
Similarity scores can also be used to select edges in a local way. The \textit{local Jaccard similarity sparsifier} (\lspar)~\cite{venu2011} includes $d^c$ edges with the highest Jaccard scores incident to each vertex locally, where $c$ is a parameter. The \localsimilarity sparsifier works similarly to \lspar, but it further ranks edges using the Jaccard score and computes $log(rank(edge))/log(deg(v))$ as the similarity score. Finally, \localsimilarity sparsifier selects edges with the highest similarity scores.

The \lspar and \localsimilarity sparsifiers are particularly useful for preserving local structure in the graph such as clustering. They can be applied to social network analysis and recommendation systems. By focusing on local similarities between vertices, these sparsifiers provide a more accurate representation of the original graph's local properties compared to other sparsifiers.

\begin{table*}[ht!]
\scriptsize
\begin{tabular}{cccccrrrc}
\textbf{Category} & \textbf{Name} & \textbf{Directed?} & \textbf{Weighted?} & \textbf{Connected?} & \multicolumn{1}{c}{\textbf{\#Nodes}} & \multicolumn{1}{c}{\textbf{\#Edges}} & \multicolumn{1}{c}{\textbf{Density}} & \textbf{source} \\ \hline \hline
\multirow{2}{*}{Social Network} & \facebook & \redcheck & \redcheck & \greencheck & 4,039 & 88,234 & 1.08E-02 & snap~\cite{snapnets} \\
 & \twitter & \greencheck & \redcheck & \redcheck & 81,306 & 1,768,149 & 2.67E-04 & snap~\cite{snapnets} \\ \hline
gene & \human & \redcheck & \greencheck & \redcheck & 14,340 & 9,041,364 & 8.79E-02 & SuiteSparse~\cite{suitesparse} \\ \hline
\multirow{2}{*}{\begin{tabular}[c]{@{}c@{}}Community \\ Network\end{tabular}} & \dblp & \redcheck & \redcheck & \greencheck & 317,080 & 1,049,866 & 2.09E-05 & snap~\cite{snapnets} \\
 & \amazon & \redcheck & \redcheck & \greencheck & 334,863 & 925,872 & 1.65E-05 & snap~\cite{snapnets} \\ \hline
communication & \enron & \redcheck & \redcheck & \redcheck & 36,692 & 183,831 & 2.73E-04 & snap~\cite{snapnets} \\ \hline
\multirow{2}{*}{collaboration} & \astroph & \redcheck & \redcheck & \redcheck & 18,772 & 198,110 & 1.12E-03 & snap~\cite{snapnets} \\
 & \hepph & \redcheck & \redcheck & \redcheck & 12,008 & 118,521 & 1.64E-03 & snap~\cite{snapnets} \\ \hline
\multirow{4}{*}{web} & \berkstan & \greencheck & \redcheck & \redcheck & 685,230 & 7,600,595 & 1.62E-05 & snap~\cite{snapnets} \\
 & \google & \greencheck & \redcheck & \redcheck & 875,713 & 5,105,039 & 6.66E-06 & snap~\cite{snapnets} \\
 & \notredame & \greencheck & \redcheck & \redcheck & 325,729 & 1,497,134 & 1.41E-05 & snap~\cite{snapnets} \\
 & \stanford & \greencheck & \redcheck & \redcheck & 281,903 & 2,312,497 & 2.91E-05 & snap~\cite{snapnets} \\ \hline
\multirow{2}{*}{GNN} & \reddit & \redcheck & \redcheck & \greencheck & 232,965 & 57,307,946 & 2.11E-03 & pyg~\cite{hamilton2017} \\
 & \proteins & \redcheck & \redcheck & \greencheck & 132,534 & 39,561,252 & 4.50E-03 & ogb~\cite{szklarczyk2018, hu2020ogb} \\ \hline \hline
\end{tabular}
\caption{Graph datasets information.}
\label{tab:datasets}
\end{table*}

\subsubsection{Effective Resistance (\er) Sparsifier}\hfill\\
The concept of Effective Resistance (\er) is derived from the analogy of an electrical circuit and applied to a graph. In this context, edges represent resistors, and the effective resistance of an edge corresponds to the potential difference generated when a unit current is introduced at one end of the edge and withdrawn from the other.

We refer readers to \cite{spielman2011} for the details of how \er is calculated. 
Once the effective resistance is calculated, a sparsified subgraph can be constructed by selecting edges with a probability proportional to their effective resistances. Notably, Spielman and Srivastava further proved that the quadratic form for Laplacian of such sparsified graphs is close to that of the original graph. Then the following inequality holds for the sparsified subgraph with high probability:
\begin{equation*} \label{eq:quadratic}
\forall \boldsymbol{x}\in \mathbb{R}^{|\mathcal{V}|} \ \ \ (1-\epsilon)\boldsymbol{x}^T\boldsymbol{L} \boldsymbol{x}\le \boldsymbol{x}^T\Tilde{\boldsymbol{L}}\boldsymbol{x} \le(1+\epsilon)\boldsymbol{x}^T\boldsymbol{L} \boldsymbol{x}
\end{equation*}
where $\Tilde{\boldsymbol{L}}$ is the Laplacian of the sparsified graph, and $\epsilon > 0$ is a small number. The insight is that \er reflects the significance of an edge. \er is a spectral sparsifier, and it aims to preserve quadratic form of the graph Laplacian. It can be applied to applications that rely on the quadratic form of graph Laplacian, for example, min-cut/max-flow.

We list the sparsifiers discussed in the section and their applicability to types of graphs, features, and time complexity in table~\ref{tab:sparsifiers}.

\subsection{Datasets}
Table~\ref{tab:datasets} lists the graph datasets used in this work, we select graphs from various categories that have different characteristics, sizes, and densities to ensure the diversity of graphs.

\section{Experimental Setup}
\subsection{Graph Preparation}
The graphs employed in this study are sourced from multiple graph dataset suites. 
We carry out essential pre-processing steps on all graphs to ensure their proper preparation for sparsifier execution and metric evaluation.
The process can be summarized as follows:
\begin{enumerate}
    \item We remove vertices with no edge incidence (i.e., isolated vertices), as they do not contribute to graph information and can induce noise in metric evaluations. Then vertices are re-indexed to be zero-based and continuous.
    \item For each directed graph, an undirected version is generated by symmetrizing each edge (i.e., adding a [dst, src] edge to the graph if it does not already exist). This ensures that sparsifiers that only operate on undirected graphs can function properly. Other sparsifiers are still applied to the original directed graphs. 
\end{enumerate}

\subsection{Graph Sparsification}
In this section, we cover additional information regarding the graph sparsifiers.
When applying sparsifiers:
\begin{enumerate}
    \item We sweep the prune rate from 0.1 to 0.9, with a step of 0.1. Some sparsifiers have a coarser prune rate granularity (e.g., \kneighbor, \lspar), and we attempt to align them with the specified prune rate. Some sparsifiers have a maximum prune rate (e.g., \localdegree, \kneighbor), so we sweep up to their maximum prune rate. Certain sparsifiers have no control over the prune rate and only support a single prune rate (e.g., \spanning, \tspanner), and we retain them as is.
    \item For non-deterministic sparsifiers, the inherent randomness in the algorithm produces different sub-graphs in each run. In such cases, we generate 10 graphs at each prune rate, measure graph metrics using the mean value, and indicate their standard deviation in the results. For deterministic sparsifiers, we generate a single graph at each prune rate.
    \item For the Effective Resistance sparsifier, since it is the only one that modifies edge weights, we consider two variants denoted as \erw and \eruw, respectively.
\end{enumerate}

\subsection{Graph Metrics}
In this section, we cover additional information regarding the measurement of sparsifiers' quality on graph metrics.

\subsubsection{Basic Metrics} \hfill\\
\textbf{Graph connectivity.}
To measure graph connectivity, we employ the source-destination pair unreachable ratio and the vertex isolated ratio.
The former represents the fraction of vertex pairs that do not have a path connecting them. The latter signifies the proportion of vertices that are isolated, meaning no edges are incident to them. Both of these ratios provide insights into the overall connectivity of a graph when assessing the effectiveness of sparsification methods.

\textbf{Degree Distribution.} 
We assess how closely the similarity of the degree distribution of the sparsified graphs and that of the original graph using the Bhattacharyya distance~\cite{bhattacharyya1946}, defined as:
\begin{equation*} \label{eq:bhattacharyya}
    B_d(P,Q) = -ln\left(\sum_{x\in\mathcal{X}}\sqrt{P(x)Q(x)}\right)
\end{equation*}
where $P$ and $Q$ are two distributions.
A value closer to 0 indicates a higher similarity in distribution. We evenly divide the discrete degree distribution into 100 bins for all graphs.

\textbf{Quadratic Form Similarity.} 
To evaluate this, we generate 100 vectors $\boldsymbol{x}$ with random entries. Next, we compute the quadratic form $\boldsymbol{x}^T \boldsymbol{L} \boldsymbol{x}$ for the original and the sparsified graphs. Then we use the mean quadratic form ratio to assess the sparsification quality.

\subsubsection{Distance Metrics} \hfill \\
\textbf{APSP and Eccentricity.} 
The computation of the All-Pair-Shortest-Path (APSP) is time-consuming for large graphs. Therefore, we resort to randomly sampling 100,000 source-destination pairs, referred to as Some-Pair-Shortest-Path (SPSP), and report the average stretch factor, which is defined as the distance ratio between the same pair in the sparsified and the original graph. We exclude pairs belonging to different communities.
Similarly, we randomly select 1000 vertices to represent the eccentricity of all vertices.

\textbf{Diameter.} 
Computing the true diameter requires performing APSP, which is impractical on large graphs. We employ an approximate diameter algorithm~\cite{dijkstra1959}. The algorithm starts with a randomly chosen source vertex, identifies a target vertex farthest from it, and iteratively repeats the process using the target vertex as the new source vertex. We validated the approximate diameter against the true diameter on small graphs and verified that they are closely aligned.
To minimize potential bias introduced by the initial source vertex selection, each graph is assessed using 10 different randomly chosen seed vertices to obtain the mean diameter.

\subsubsection{Centrality Metrics} \hfill\\
We employ the top-k precision to evaluate the quality of centrality metrics. First, vertices are ranked according to their centrality scores. Then the top-k vertices in the sparsified graphs are compared with those in the full graph. The proportion of overlapping vertices is referred to as the top-k precision. In this paper, we set k to 100 because typically only a small subset of vertices in graphs are critical and accurately ranking them is more important.

\textbf{Betweenness Centrality.} 
Actual betweenness centrality calculation also requires computing APSP. In this paper, we adopt an approximate betweenness centrality algorithm proposed by Geisberger \textit{et al.}~\cite{robert2008}. The algorithm is sampling-based, and a higher sampling number achieves better estimation quality. We use a sampling number of 500 and compare it with exact betweenness on a set of small graphs, confirming the results are closely aligned.

\subsubsection{Application-level Metrics} \label{sec:highlevel_setup} \hfill\\
\textbf{Min-cut/Max-flow.} We randomly sample 100,000 src-dst pairs and measure the min-cut/max-flow on both the original and sparsified graphs. Then we use the mean stretch factor between the sparsified and the original graph to evaluate the sparsification quality. 

\textbf{GNN.} For GNNs, we evaluate two models: GraphSAGE and ClusterGCN. The quality is measured in test accuracy or Area Under the Receiver Operating Characteristics~\cite{fawcett2006} (AUROC). 
AUROC ranges from 0.5 to 1. A higher accuracy or AUROC indicates better GNN performance. For both GNN models, we train the network with sparsified graph and test on the full graph, because 1) training is the most time-consuming part and is the most meaningful to apply sparsification, 2) testing on the full graph reveals how well the sparsified graph captures full graph's characteristics.

\subsection{Software Framework}
Our software evaluation framework integrates several open-source libraries and our custom implementations. We use \texttt{NetworKit}\cite{networkitpaper} for multiple sparsifiers and \texttt{Laplacians.jl}\cite{ER_github} for the effective resistance sparsifier. We also implemented the \kneighbor, \rankdegree, \lspar, and \tspanner algorithms.

For the evaluation metrics, we employ both \texttt{NetworKit}~\cite{networkitpaper} and \texttt{graph-tool}~\cite{graph-tool_2014} for implementations of several discussed distance, centrality, clustering, and min-cut/max-flow metrics. We use \texttt{PyG}~\cite{pyg} to implement the graph neural networks. Additionally, we implemented degree distribution and quadratic form evaluation.

The framework is open-sourced and extendable to incorporate more sparsification algorithms and graph metrics.

\subsection{Hardware Platform}
The experiments in this paper are performed on a server with an Intel Xeon Platinum 8380 CPU, with 1 TB memory. The graph neural networks run on an Nvidia A40 GPU with 48 GB memory.

\section{Results}

In this section, we evaluate the impact of various sparsifiers on the quality of graph metrics at different prune rates. We perform comprehensive experiments on all sparsifiers, graph metrics, and datasets discussed in this paper. Due to the extensive nature of the experiments we conducted (over 30,000 data points), we can only show a subset of performance results in the figures. The full results are available with the artifact. We adhere to the following rules to present the results without bias: (1) for readability, we only show a representative subset of sparsifiers for each graph metric, including those that perform well or poorly and those that yield interesting outcomes; (2) we always include \random as it serves as a naive sparsifier for comparison; (3) we select at least one representative graph for each graph metric and discuss any discrepancies observed in other graphs. We then compare sparsification times and briefly discuss the overhead associated with sparsification. Finally, We summarize the results and provide insights. 

\subsection{Basic Metrics}
\begin{figure}[h!]
\begin{subfigure}[b]{0.45\textwidth}
    \includegraphics[width=0.8\linewidth, center, trim={0 1cm -3cm 2cm}]{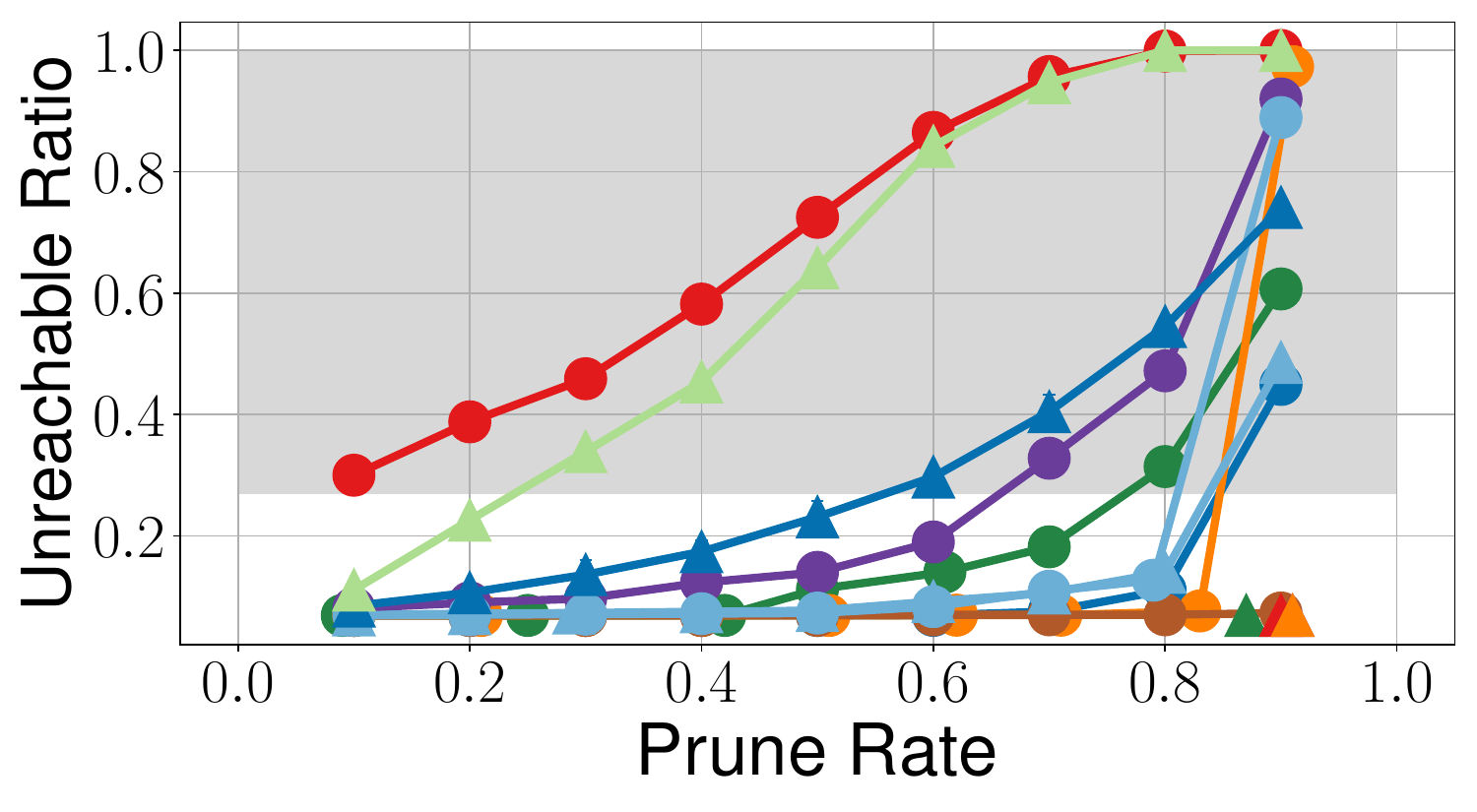}
    \caption{Pair Unreachable Ratio}
    \label{fig:spsp_unreachable}
\end{subfigure}
\begin{subfigure}[b]{0.5\textwidth}
    \centering
    \includegraphics[width=0.75\linewidth, trim={0 1cm 0 0}]{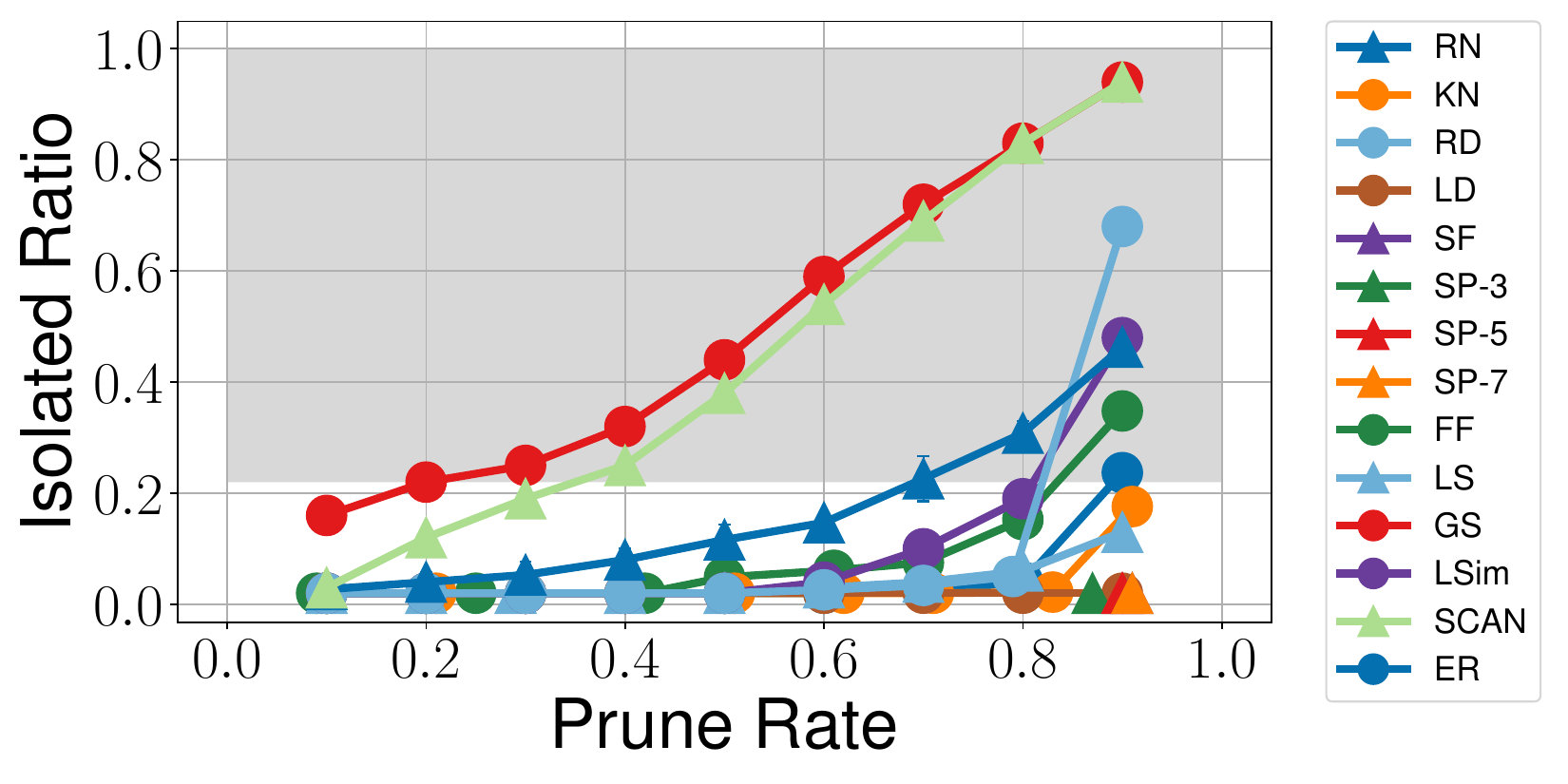}
    \caption{Vertex Isolated Ratio}
    \label{fig:eccentricity_isolated}
\end{subfigure}
    \caption{Graph Connectivity on \astroph.}
    \label{fig:graph_connectivity}
\end{figure}
Figures~\ref{fig:spsp_unreachable} and \ref{fig:eccentricity_isolated} show the source-destination pair unreachable ratio and vertex isolated ratio, respectively. As the prune rate increases, the graph becomes more disconnected, leading to an increase in isolated vertices. \kneighbor excels at preserving graph connectivity because it ensures that each vertex retains at least $k$ edges. Two local sparsifiers, \localdegree and \localsimilarity, also show strong performance since they both select edges to maintain locally, guaranteeing at least one edge for each vertex. \er performs well by retaining high-resistance edges, which are the low-redundancy edges crucial for maintaining graph connectivity. \spanning and \tspanners preserve the same level of connectivity as the original graph, as ensured by the algorithms. \random does not effectively preserve graph connectivity because it does not attempt to maintain edges critical for connectivity. \gspar and \scan retain edges connecting similar vertices on a global scale, and these edges are often intra-community edges that are not crucial for preserving connectivity, resulting in the poorest performance. The acceptable unreachable/isolated ratio can be customized according to specific applications. In this paper, we consider an increase of 20\% or more in the unreachable/isolated ratio compared to the original graph as excessive (shown as the grey area in Figures~\ref{fig:spsp_unreachable} and \ref{fig:eccentricity_isolated}).

\begin{figure}[h!]
    \centering
    \includegraphics[width=0.7\linewidth, trim={0 0.5cm 0 0cm}, clip]{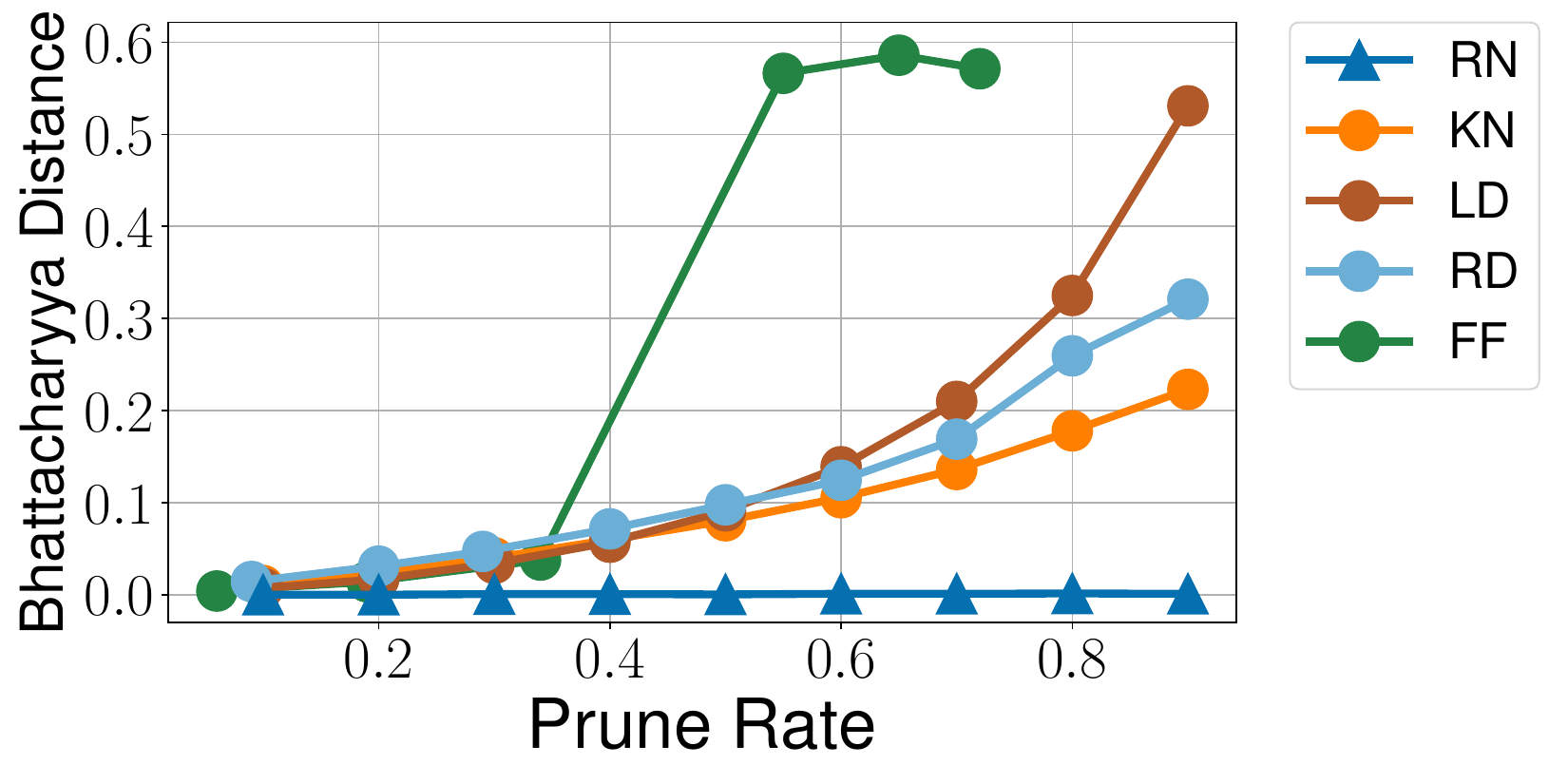}
    \caption{Degree distribution comparison on \proteins. Lower is better. \random performs the best, \localdegree and \forestfire do not do well in preserving degree distribution.}
    \label{fig:degreedistribution}
\end{figure}

\textbf{Degree Distribution.} 
Figure~\ref{fig:degreedistribution} illustrates the degree distribution on \proteins. A lower Bhattacharyya distance signifies a more similar degree distribution to the original graph. \random demonstrates the best performance in preserving the degree distribution. This is due to \random treats all edges without bias, thus maintaining the same proportion of edges for all vertices and keeping a similar degree distribution. Given that most graphs exhibit a power-law degree distribution, some sparsifiers struggle to preserve degree distribution. For instance, \localdegree and \rankdegree tend to retain edges connected to high-degree vertices. Conversely, \kneighbor maintains up to \textit{K} edges for all vertices, eliminating surplus edges from high-degree vertices. These biases negatively impact the preservation of the degree distribution. Among all sparsifiers, \random consistently performs well across all graphs, while \localdegree, \rankdegree, \kneighbor, and \forestfire under-perform on most graphs. The performance of other sparsifiers moderately fluctuates across graphs due to different graph characteristics.

\begin{figure}[h!]
    \centering
    \includegraphics[width=0.7\linewidth, trim={0 2cm 0 2cm}]{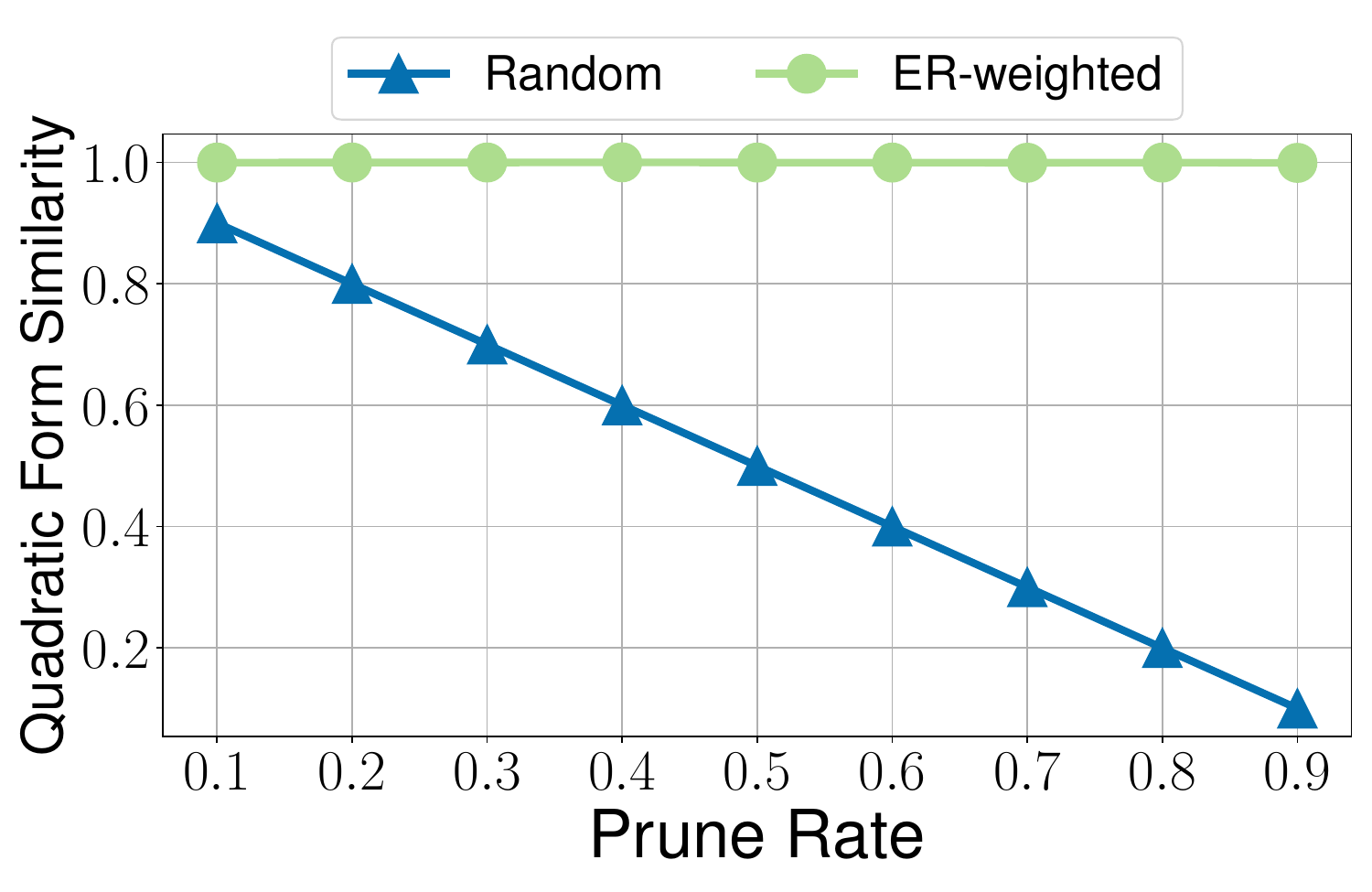}
    \caption{Laplacian quadratic form comparison of different sparsifiers on \amazon. Closer to 1 is better. \erw performs the best. \random and other sparsifiers do not preserve Laplacian quadratic form.}
    \label{fig:quadratic}
\end{figure}

\textbf{Laplacian Quadratic Form.} 
Figure~\ref{fig:quadratic} displays the Laplacian quadratic form similarity on \amazon. A value closer to 1 indicates better quality. From the figure, \erw emerges as the clear winner. This is because the Laplacian quadratic form is the specific attribute \erw is designed to preserve. Note that only \erw possesses this property. \eruw, along with other sparsifiers, exhibits no capability to preserve Laplacian quadratic form similarity at all, and they show the same pattern as \random. The pattern observed on \amazon is consistent across other undirected graphs. For directed graphs (not shown due to space limit), the Laplacian quadratic form ratio for \erw is no longer guaranteed to be close to 1, this is because the symmetrization process deviates the graph's spectral property from that of the original directed graph. However, \erw still maintains a constant ratio and offers a better guarantee than other sparsifiers.

\subsection{Distance Metrics} \label{sec:result_distance_metrics}
\begin{figure*}[h!]
\begin{subfigure}[b]{0.3\textwidth}
    \centering
    \includegraphics[width=\linewidth, trim={0 1.5cm 0 2cm}]{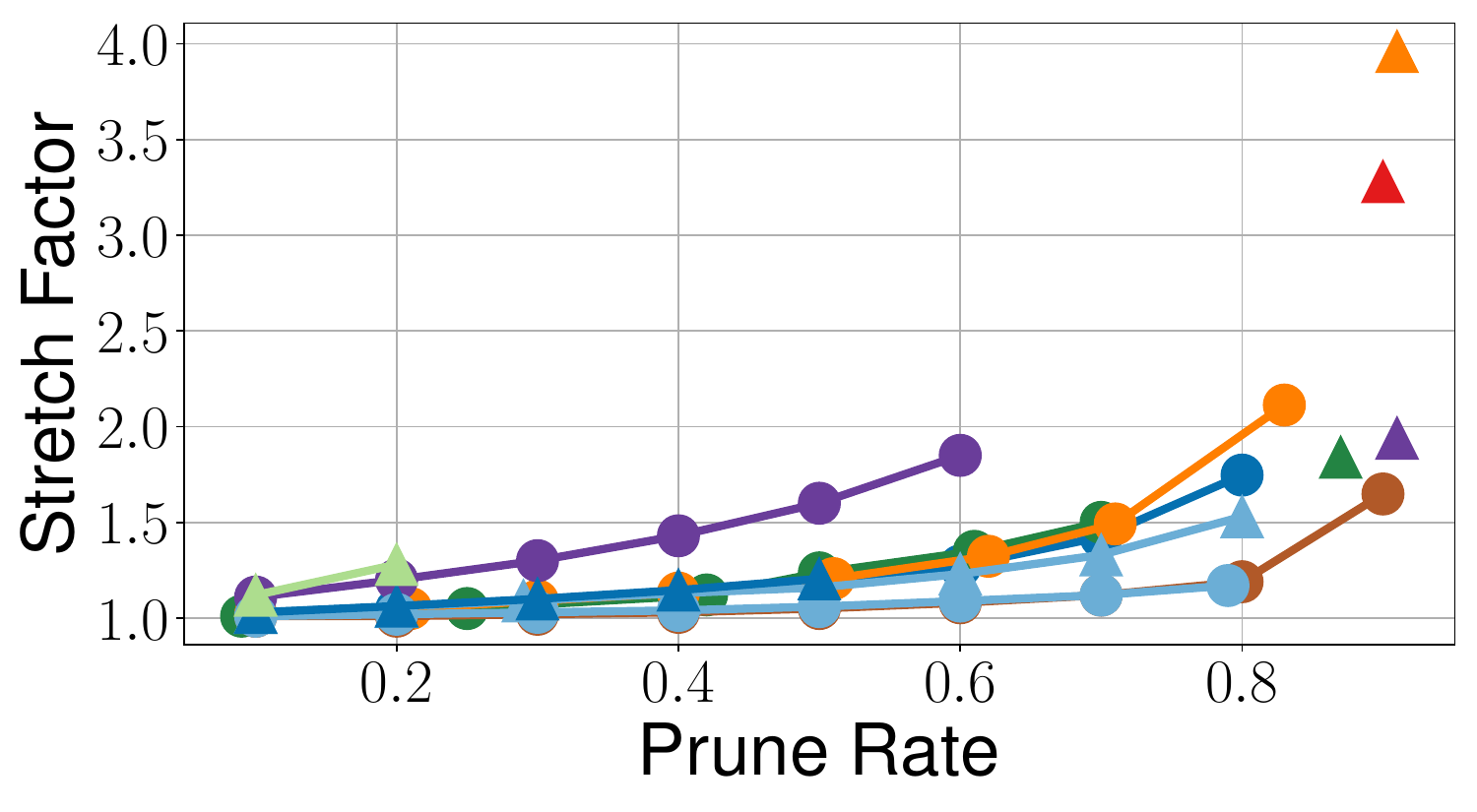}
    \caption{Adjusted SPSP Stretch Factor}
    \label{fig:adjusted_spsp_stretch_factor}
\end{subfigure}
\begin{subfigure}[b]{0.3\textwidth}
    \centering
    \includegraphics[width=\linewidth, trim={0 1.5cm 0 2cm}]{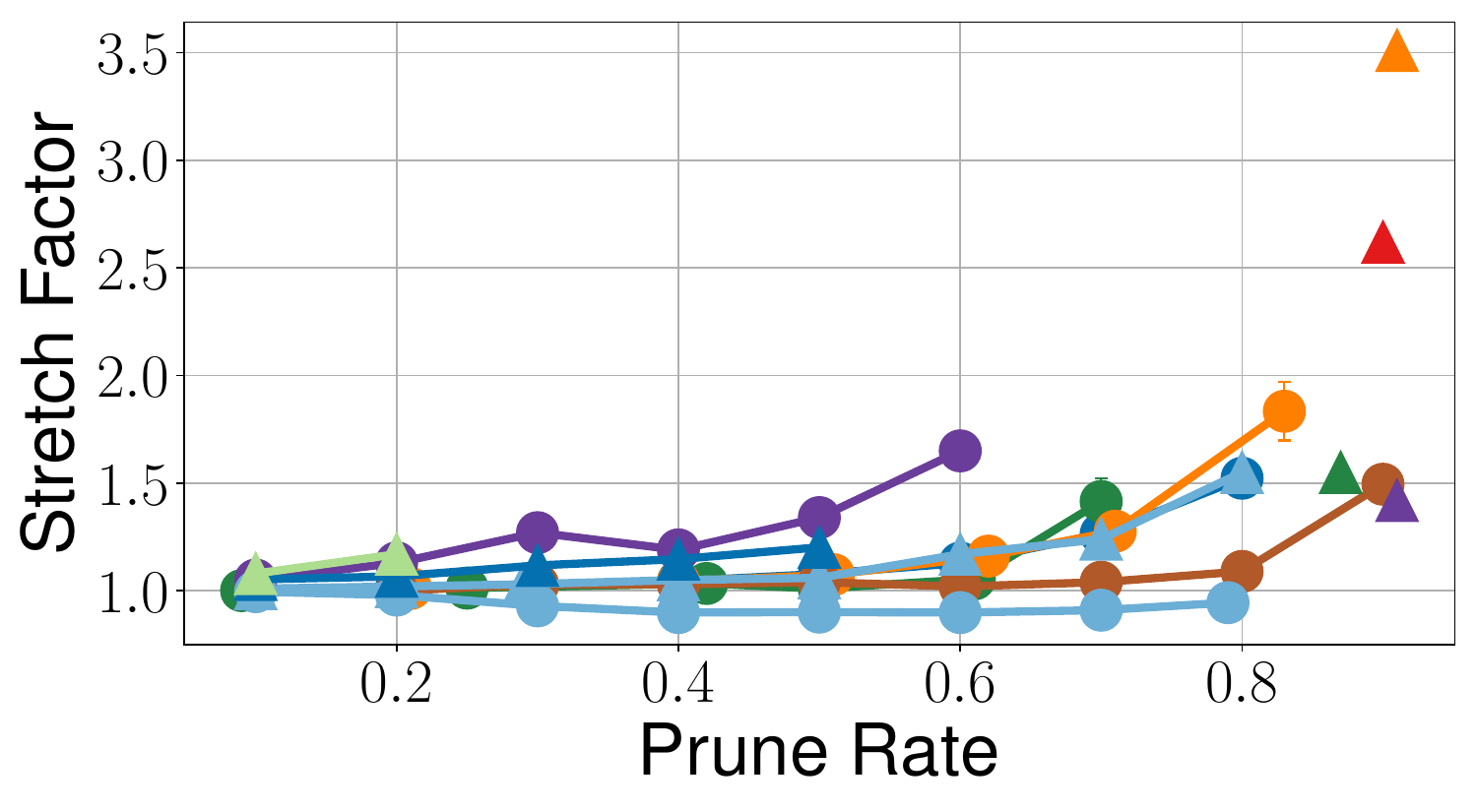}
    \caption{Adjusted Eccentricity Stretch Factor}
    \label{fig:adjusted_eccentricity_stretch_factor}
\end{subfigure}
\begin{subfigure}[b]{0.335\textwidth}
    \centering
    \includegraphics[width=\linewidth, trim={0 1.5cm 0 2cm}]{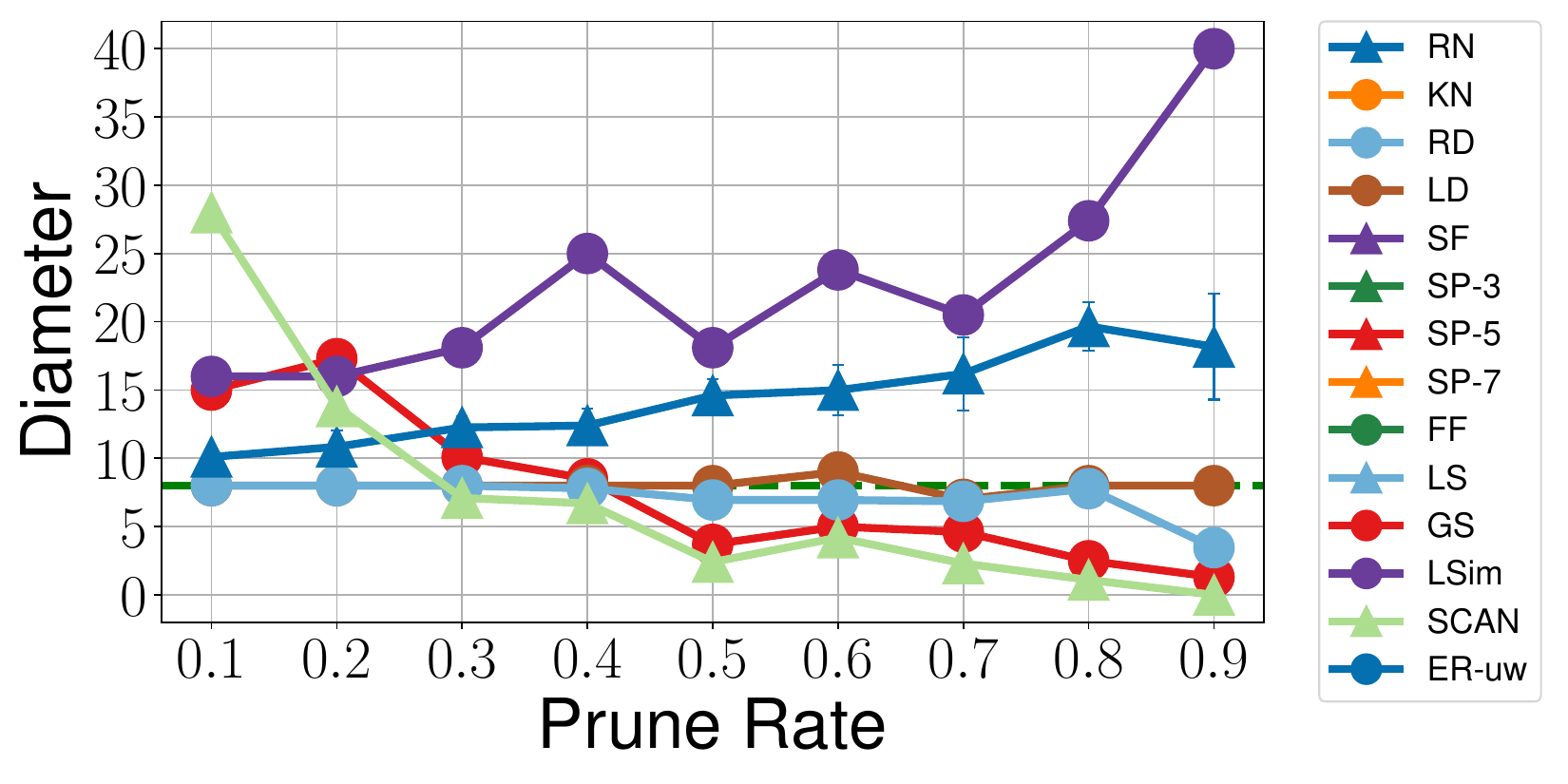}
    \caption{Diameter}
    \label{fig:diameter}
\end{subfigure}
    \caption{(a) Adjusted SPSP stretch factor of sparsifiers on \astroph with the constraint of acceptable pair unreachable ratio. (b) Adjusted eccentricity stretch factor of sparsifiers on \astroph with the constraint of acceptable vertex isolated ratio. (c) Diameter comparison on \facebook. For stretch factor, closer to 1 is better. For graph diameter, closer to ground truth (green line) is better. \rankdegree and \localdegree have the best performance. \gspar and \scan do not perform well.}
    \label{fig:SPSP}
\end{figure*}

\textbf{SPSP.} A practical sparsifier should keep the mean stretch factor close to 1 while keeping the unreachable ratio relatively low. Figure~\ref{fig:adjusted_spsp_stretch_factor} shows the mean stretch factor of 100,000 sampled source-destination pairs, with the constraint that the unreachable ratio is <20\% over that in the original graph (white area in figure~\ref{fig:spsp_unreachable}). This allows for a comparison of the mean stretch factor without a significant increase in the number of unreachable pairs. \localdegree and \rankdegree demonstrate the best performance in preserving distances while maintaining a low unreachable ratio. This is because both of them are biased towards preserving edges of high-degree vertices, which are typically hub vertices in the graph and often lie along many shortest paths.
\lspar, \eruw, \forestfire, and \kneighbor also exhibit strong performance due to their ability to maintain graph connectivity. Conversely, \gspar and \scan perform poorly as they rapidly increase the unreachable ratio and have a higher stretch factor. Although \spanning and \tspanners have a relatively high stretch factor, they guarantee the connectivity of the original graph, allowing them to maintain the unreachable ratio. \tspanners fulfill the guarantee that the stretch factor is at most t but empirically show a higher mean stretch factor than \localdegree. \tspanners is useful when connectivity is paramount and a slightly higher stretch factor is tolerable.

\textbf{Eccentricity.}
Figure~\ref{fig:adjusted_eccentricity_stretch_factor} presents the performance of sparsifiers with the vertex isolation ratio is <20\% higher than that in the original graph (white area in figure~\ref{fig:eccentricity_isolated}). 
\localdegree and \rankdegree perform the best in preserving eccentricity while keeping the unreachable ratio low. \lspar, \eruw, \forestfire, and \kneighbor also show strong performance due to their ability to maintain graph connectivity. \gspar and \scan perform poorly compared to other sparsifiers.
\spanning and \tspanners have a relatively high stretch factor but guarantee the graph connectivity. Additionally, \tspanners provide a theoretical upper bound on the stretch factor, making them suitable for certain scenarios.

\textbf{Diameter.} 
Figure~\ref{fig:diameter} presents the diameters of various sparsified graphs at various prune rates. The green dashed line (8) indicates the diameter measured on the full graph as ground truth. We observe that \localdegree and \rankdegree perform the best, consistent with their strong performance in preserving distance. \gspar, \scan, and \localsimilarity perform poorly compared to other sparsifiers.

In general, distance-related metrics are consistent across graphs. Some graphs (e.g., \amazon) have a lower average degree, causing the unreachable ratio or vertex isolation ratio to increase more quickly than in other graphs. \localdegree and \rankdegree consistently demonstrate the best performance for all distance-related metrics; however, \localdegree more effectively maintains the connectivity. \gspar and \scan always under-perform because they both tend to keep intra-community edges. This leads to a more disconnected graph and a high unreachable/isolation ratio.

\subsection{Centrality Metrics}
\begin{figure}[h!]
\begin{subfigure}[b]{0.5\textwidth}
    \centering
    \includegraphics[width=0.7\linewidth, trim={0 0.5cm 0 0cm}, clip]{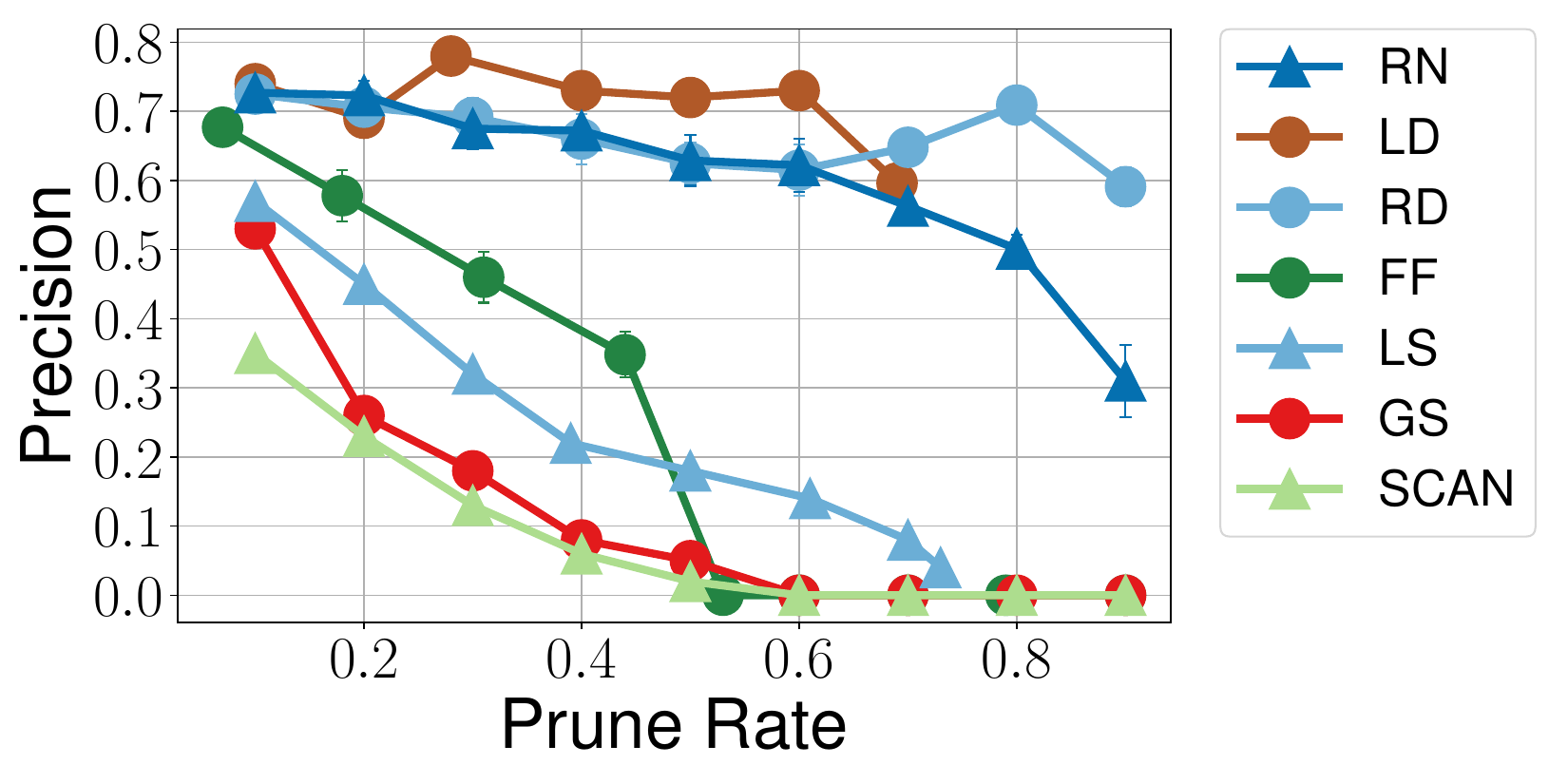}
    \caption{Betweenness Centrality}
    \label{fig:betweenness}
\end{subfigure}
\begin{subfigure}[b]{0.5\textwidth}
    \centering
    \includegraphics[width=0.7\linewidth, trim={0 0 0 0cm}, clip]{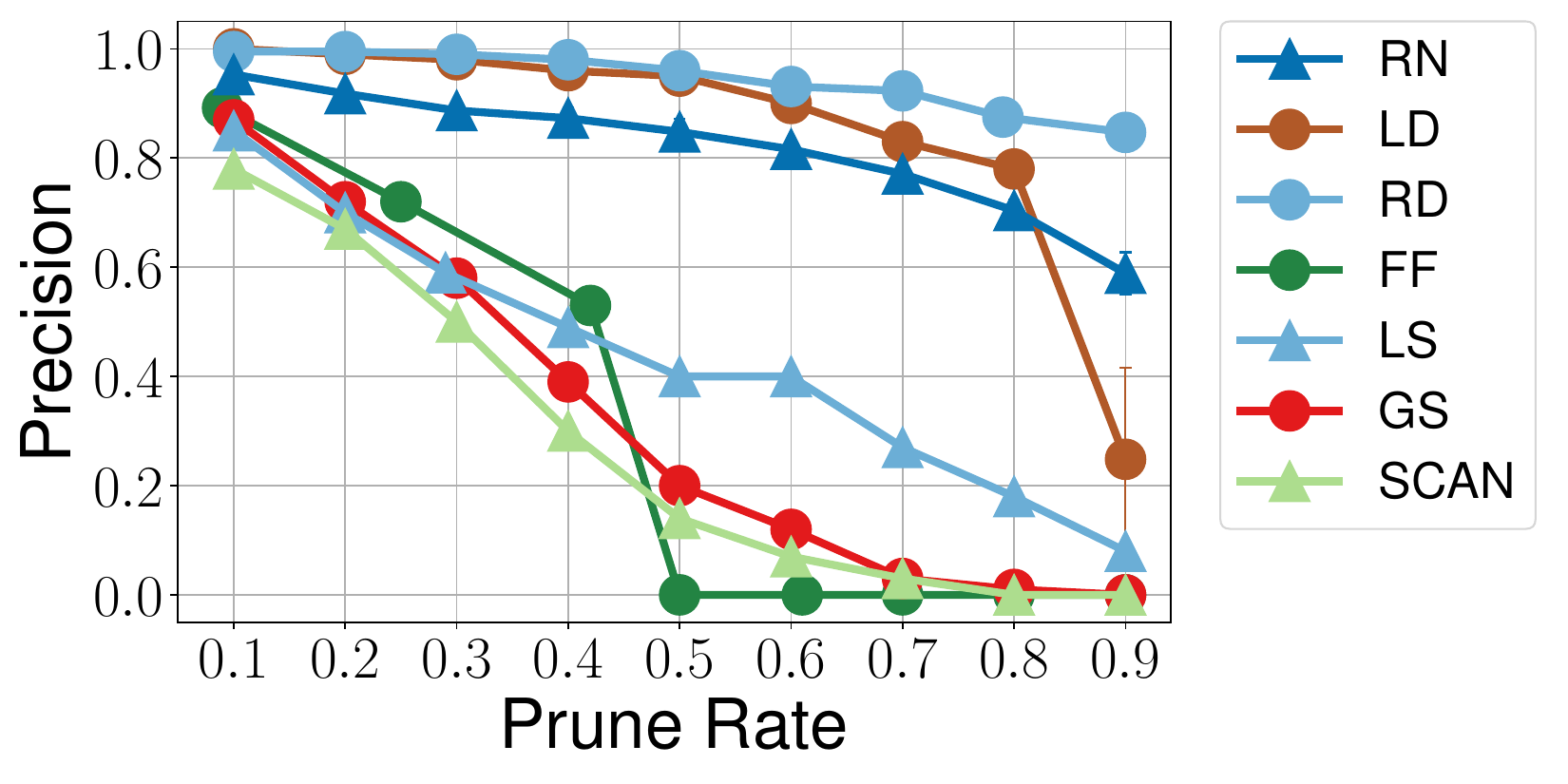}
    \caption{Closeness Centrality}
    \label{fig:closeness}
\end{subfigure}
    \caption{Top-100 precision for Betweenness and Closeness centrality. Higher is better. (a) Betweenness centrality on \dblp. (b) Closeness centrality on \astroph. \localdegree, \rankdegree, and \random have the best performance. \lspar, \gspar, \scan, and \forestfire do not perform well.}
    \label{fig:betweeness_and_closeness}
\end{figure}

\textbf{Betweenness and Closeness Centrality.} 
Figure~\ref{fig:betweenness} and \ref{fig:closeness} display the top-100 precision of betweenness centrality on \dblp and closeness centrality on \astroph. \localdegree and \rankdegree exhibit the best performance. This is because the top-scored vertices are typically hub vertices, and as explained in \S~\ref{sec:result_distance_metrics}, both \localdegree and \rankdegree preserve edges incident to high-degree vertices, thus maintaining the betweenness and closeness ranking of hub vertices. \random uniformly samples edges without bias, and preserves the relative ranking to some extent. \gspar and \scan do not perform well as they aggressively disconnect graphs. We consistently observe \localdegree, \rankdegree, and \random perform well, and \gspar and \scan perform poorly across graphs.

\begin{figure}[h!]
    \centering
    \includegraphics[width=0.7\linewidth, trim={0 0 0 0}, clip]{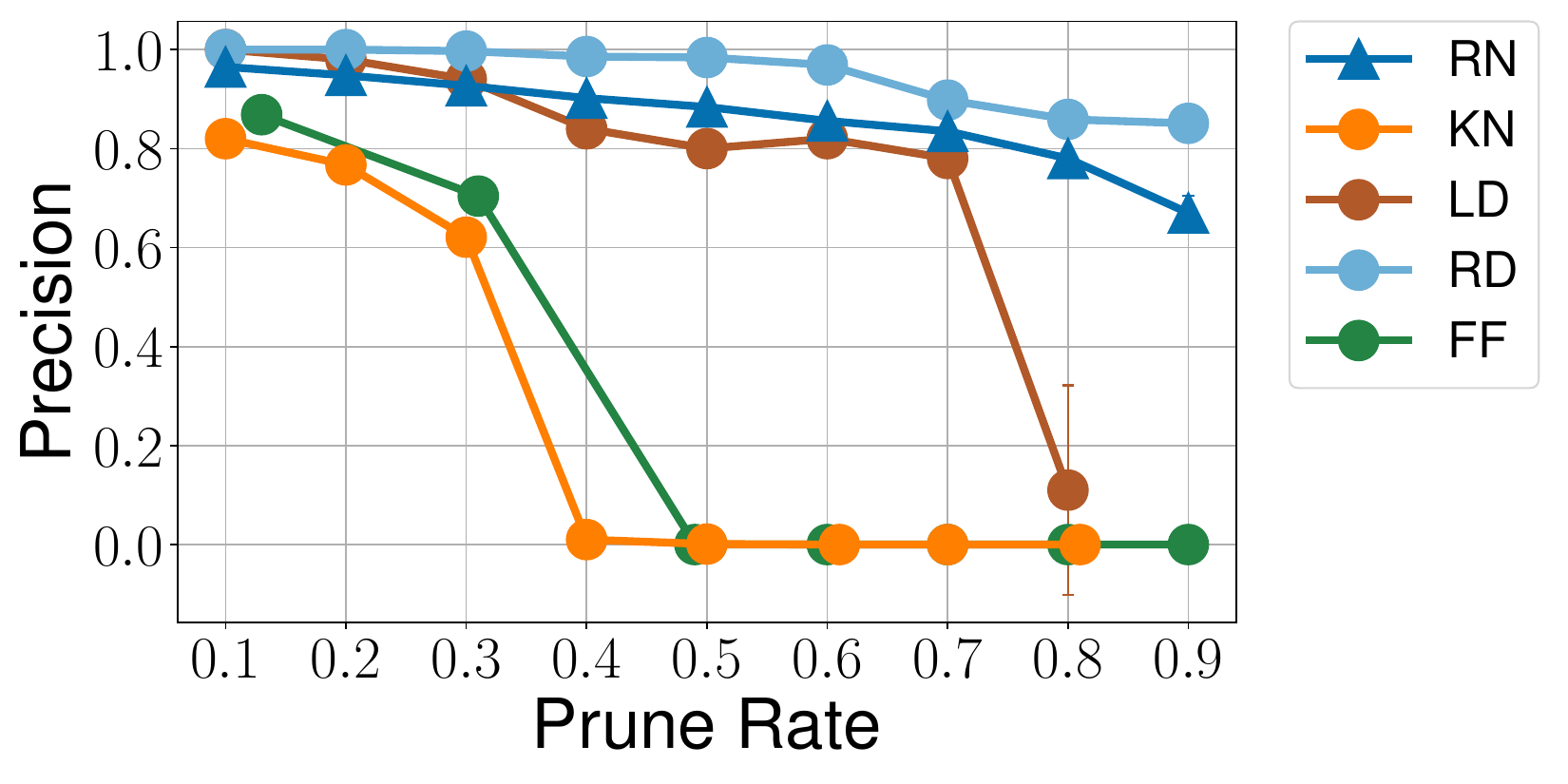}
    \caption{Eigenvector centrality top-100 precision comparison on \enron. Higher is better. \rankdegree and \random have the best performance. \forestfire and \kneighbor do not perform well.}
    \label{fig:eigenvector}
\end{figure}

\textbf{Eigenvector Centrality.} Figure~\ref{fig:eigenvector} presents the top-100 precision of eigenvector centrality on \enron. \rankdegree achieves the best performance because it retains edges connected to high-degree vertices. Although eigenvector centrality is not directly linked to degree, high-degree vertices have a higher probability of being directly or indirectly (via n-hop neighbors) connected to important vertices. In comparison, \localdegree performs worse than \rankdegree since it only considers the degree of immediate neighbors and may disconnect vertices from vital vertices located more than 1-hop away. \random shows strong performance due to its unbiased nature, which helps preserve relative ranking. Both \forestfire and \kneighbor under-perform in preserving eigenvector centrality.

\begin{figure}[h!]
    \centering
    \includegraphics[width=0.7\linewidth, trim={0 0.5cm 0 0}, clip]{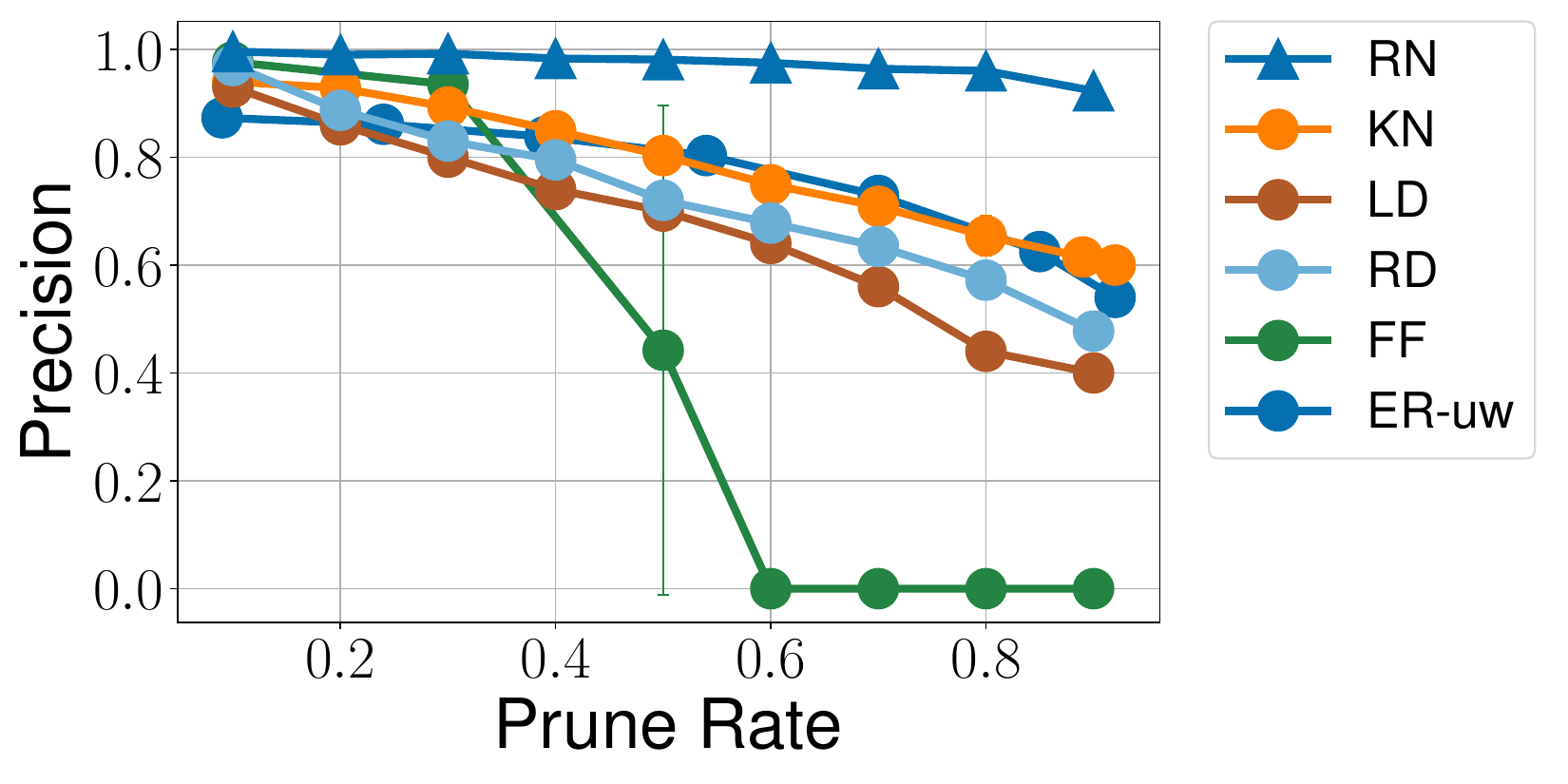}
    \caption{Katz centrality top-100 precision comparison of different sparsifiers on \twitter. Higher is better. \random has the best performance. \forestfire does not perform well.}
    \label{fig:katz}
\end{figure}

\textbf{Katz Centrality.} 
Figure~\ref{fig:katz} illustrates the top-100 precision of Katz centrality on \twitter. \random demonstrates the most effective performance. This is due to that \random proportionally maintains the number of edges relative to the original degree for all vertices. Thus, the graph's hop structure closely resembles its original state. Empirically, \kneighbor and \eruw also exhibit strong performance. \localdegree and \rankdegree do not perform well since they solely focus on degree, thereby only accounting for immediate neighbors. Therefore, vertices with low-degree immediate neighbors but high k-hop (k>1) neighbors are severely penalized. Minor fluctuations in sparsifiers' relative performance on certain graphs can be attributed to the variation in the attenuation factor $\alpha$. Overall, the performance is consistent across graphs.

In summary, \localdegree, \rankdegree, and \random consistently excel in centrality-related metrics. This is because \localdegree and \rankdegree retain edges connected to hub vertices, and centrality metrics seek important vertices in the graph, which often correspond to hub vertices. Conversely, \random maintains edges without bias, thus effectively preserving the relative vertex ranking.

\subsection{Clustering Metrics}
\begin{figure}[h!]
    \centering
    \includegraphics[width=0.7\linewidth, trim={0 0 0 0}, clip]{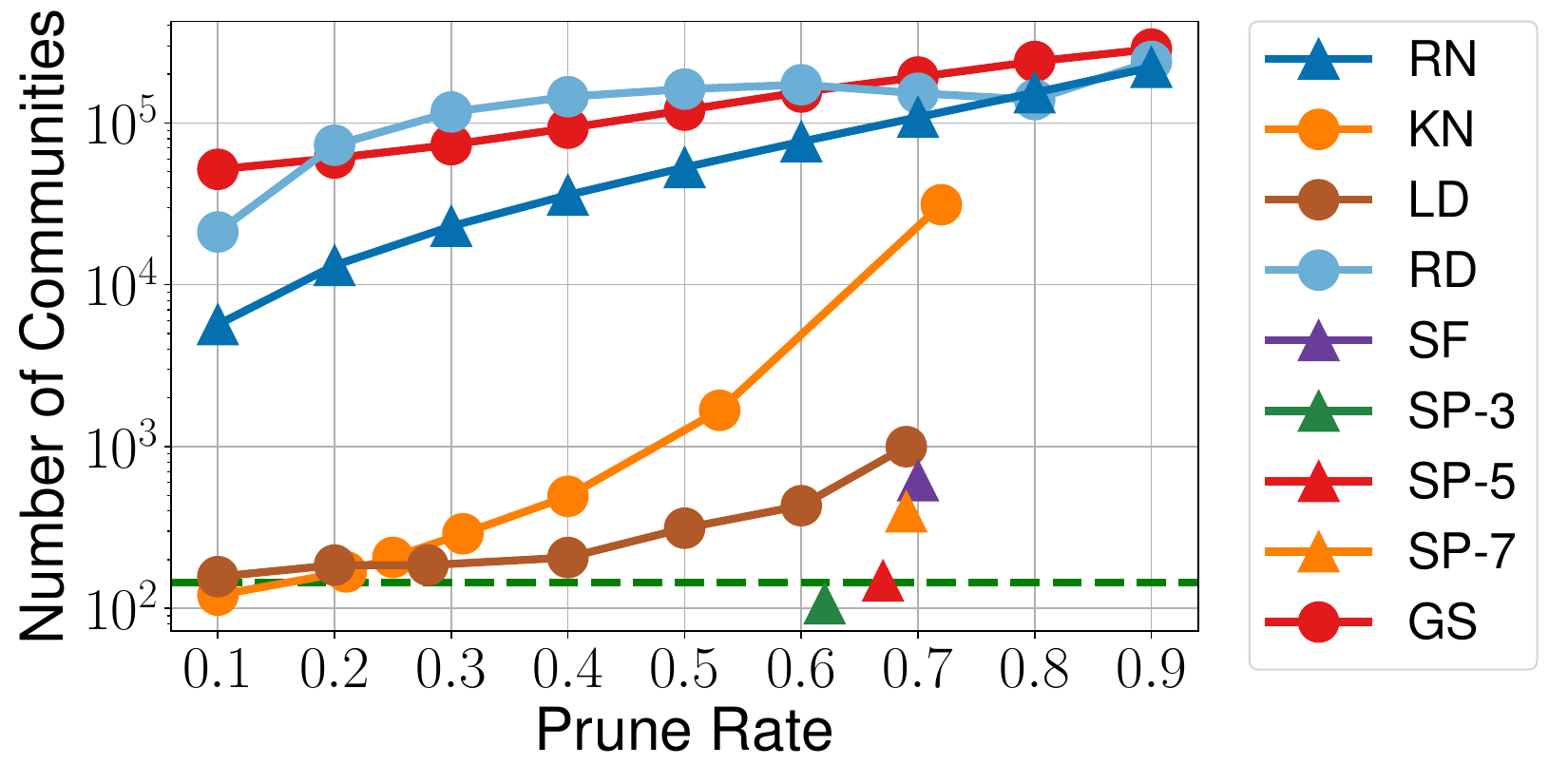}
    \caption{Number of communities comparison on \dblp. Closer to the green line is better. \localdegree, \spanning, and \tspanners have the best performance. \gspar, \rankdegree, and \random do not perform well.}
    \label{fig:num_community}
\end{figure}
\textbf{Number of Communities.} We employ the widely recognized Louvain method~\cite{Blondel2008} for community detection, assuming the number of communities is unknown, and use the number detected in the original graph as the ground truth. Figure~\ref{fig:num_community} presents a comparison of community numbers on \dblp, with the green dashed line representing the ground truth; the closer to it, the better. As the prune rate increases, the graph becomes increasingly disconnected, and the number of communities consistently rises. \localdegree and \kneighbor excel in maintaining the community number relatively close to the ground truth because it preserves the connectivity. \spanning and \tspanners also demonstrate strong performance, surpassing \localdegree at equivalent prune rates, as they ensure connectivity remains identical to the original graph. Unlike \localdegree, \rankdegree struggles to preserve the community number because it sparsifies globally without providing guarantees on connectivity preservation. In various graphs, \localdegree, \spanning, and \tspanners consistently outperform.

\begin{figure}[h!]
\begin{subfigure}[b]{0.5\textwidth}
    \centering
    \includegraphics[width=0.7\linewidth, trim={0 0 0 0}, clip]{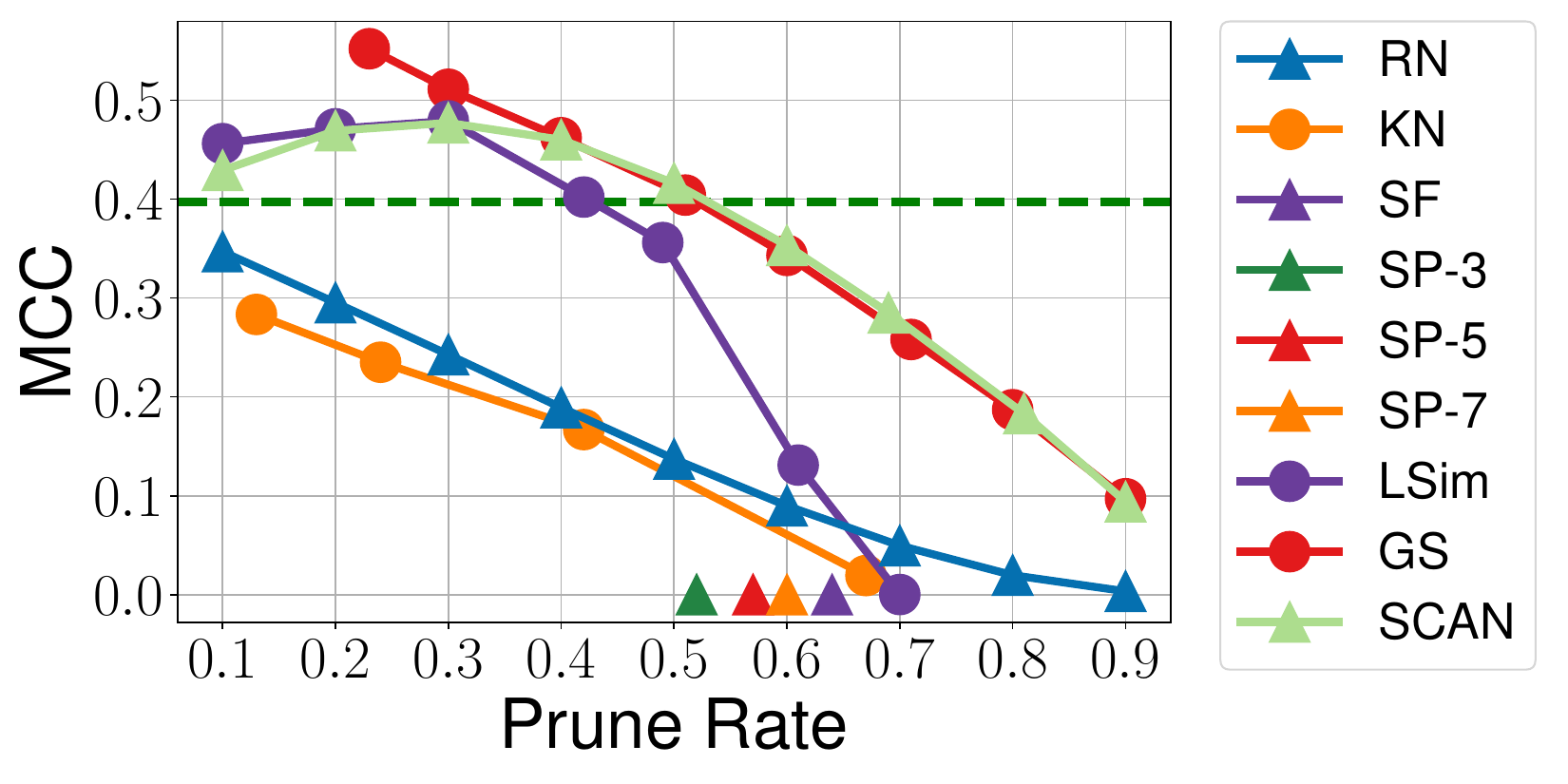}
    \caption{Mean clustering coefficient.}
    \label{fig:mcc}
\end{subfigure}
\begin{subfigure}[b]{0.5\textwidth}
    \centering
    \includegraphics[width=0.7\linewidth, trim={0 0 0 0}, clip]{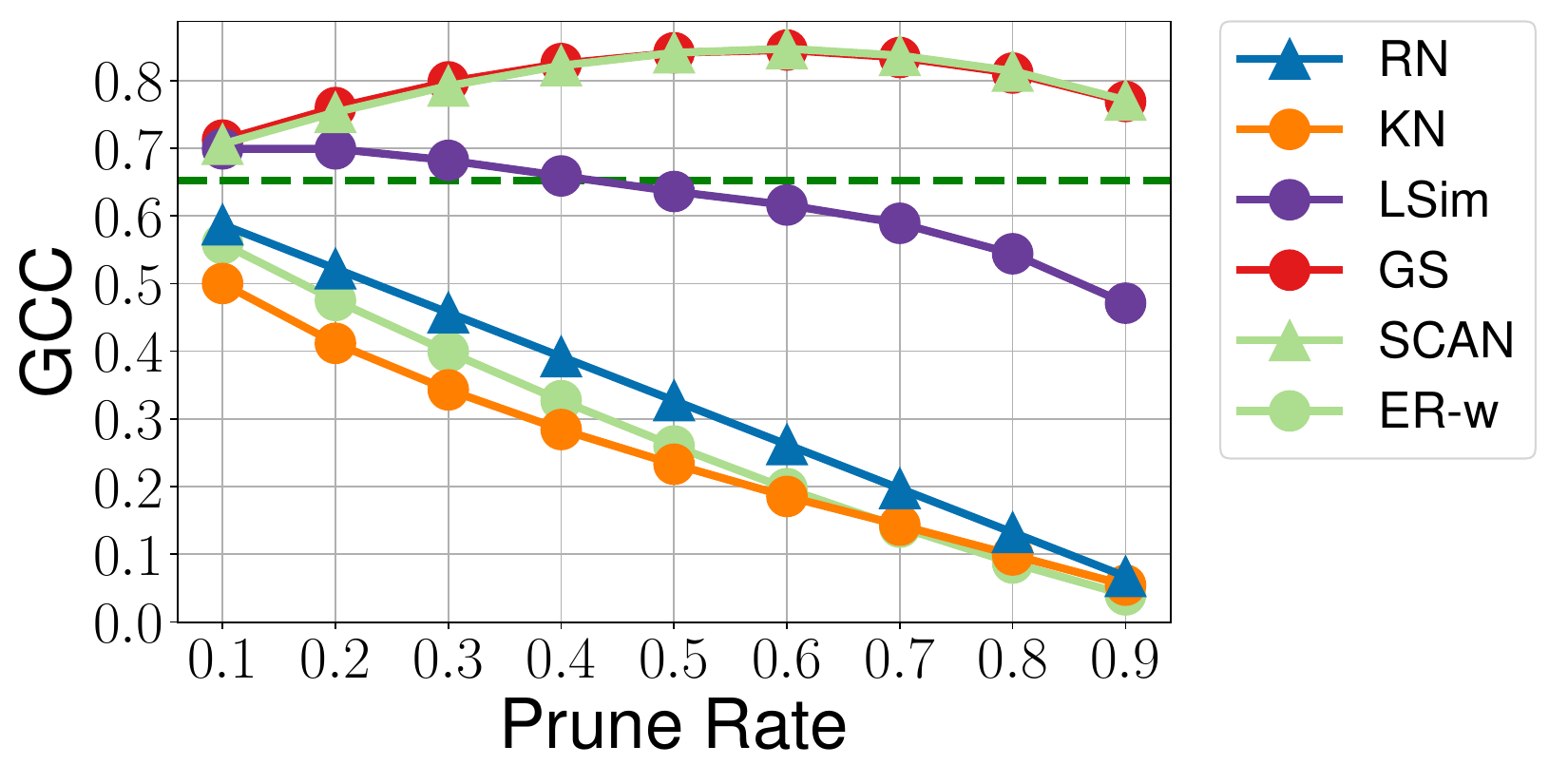}
    \caption{Global clustering coefficient.}
    \label{fig:gcc}
\end{subfigure}
    \caption{Clustering coefficients comparison. Closer to the green line is better. (a) shows the MCC on \amazon (b) shows the GCC on \human. No sparsifier is effective in preserving the clustering coefficient.}
    \label{fig:clusteringcoeff}
\end{figure}

\textbf{Clustering Coefficient.} Figure~\ref{fig:clusteringcoeff} compares clustering coefficients on \amazon and \human. We use the mean clustering coefficient (MCC) to evaluate the local clustering coefficient (LCC), as it represents the average LCC of all vertices. The green dashed lines indicate the MCC and GCC of the original graph. Generally, most sparsifiers exhibit decreasing MCC and GCC as the prune rate rises, with only \localsimilarity, \scan, and \gspar exhibiting slight increases in MCC at lower prune rates. None of the sparsifiers demonstrate outstanding performance in preserving MCC and GCC, as they all degrade linearly with respect to the prune rate. \spanning and \tspanners consistently have an MCC of 0 due to the absence of loops in the graph. Clustering coefficient results vary across different graphs, with graph categories and directedness significantly impacting sparsifier performance. Overall, no sparsifier proves effective in preserving clustering coefficients.

\begin{figure}[h!]
    \centering
    \includegraphics[width=0.7\linewidth, trim={0 0 0 0}, clip]{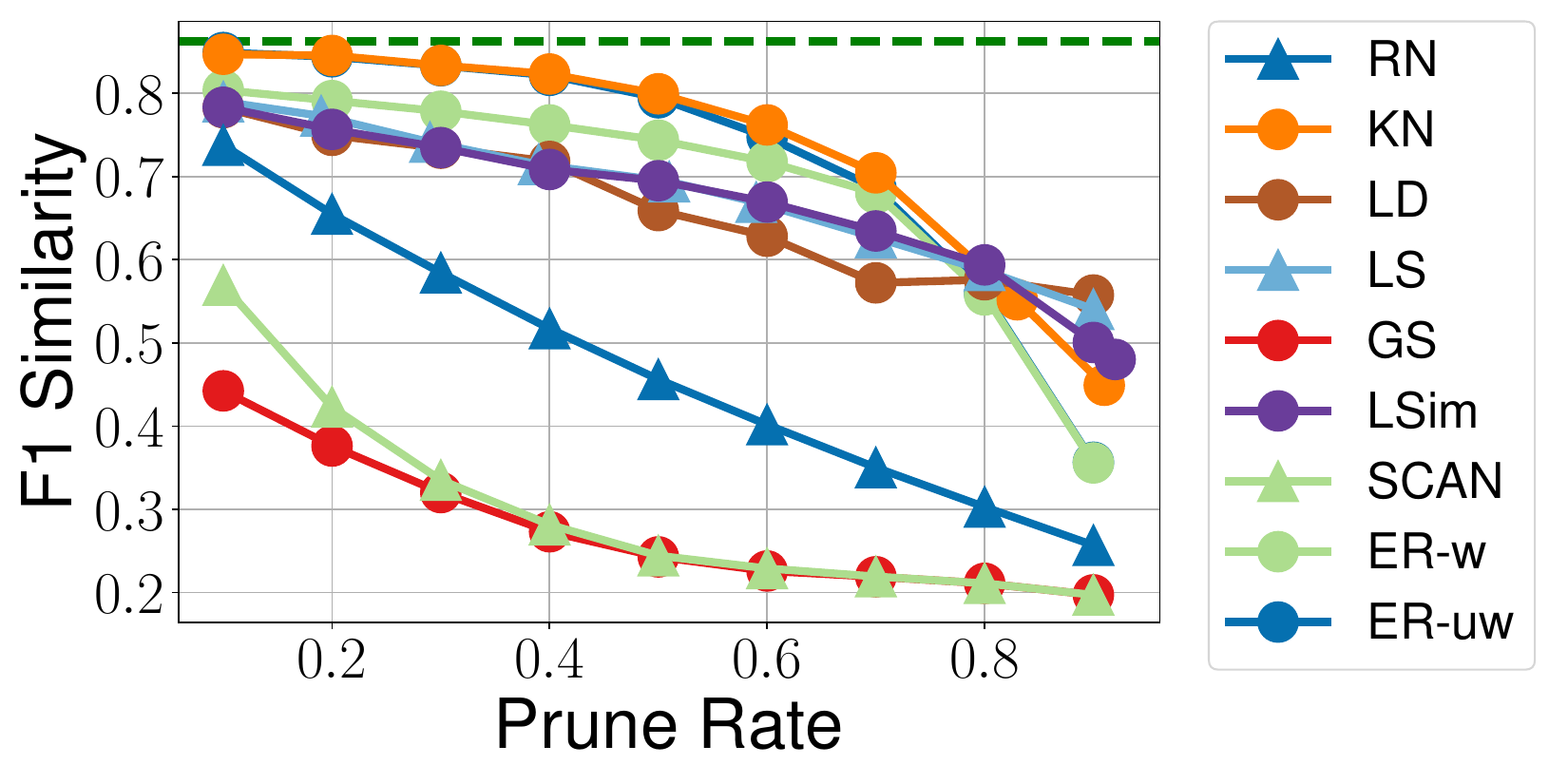}
    \caption{Clustering F1 similarity comparison of different sparsifiers on \hepph. Higher is better. \eruw, \erw, \kneighbor, \localdegree, \lspar, and \localsimilarity perform the best. \scan and \gspar underperform.}
    \label{fig:clusteringf1}
\end{figure}

\textbf{Clustering F1 Similarity.} Relying solely on the number of communities to evaluate clustering quality is insufficient, therefore, we employ the clustering F1 score to measure clustering similarity (see \S~\ref{sec:cluster_metrics}). Figure~\ref{fig:clusteringf1} shows the clustering F1 similarity comparison on \hepph, with F1 similarity ranging from 0 (worst) to 1 (best). The green dashed line represents the clustering F1 similarity when applying clustering algorithms twice on the original graph; it is not 1 due to the inherent randomness in the clustering algorithm.
For all sparsifiers, F1 similarity decreases as the prune rate increases. \kneighbor exhibits the best overall performance, while \localsimilarity, \localdegree, and \lspar also demonstrate strong results. These sparsifiers share a focus on local edges, and locally similar vertices more likely to belong to the same community. Empirically, we also observe that \erw and \eruw perform well, potentially due to \er's preservation of low-redundant edges, which are often crucial in clustering algorithms. In contrast, \gspar and \scan display poor performance in preserving clustering similarity. Across graphs, \kneighbor, \localdegree, \localsimilarity, \lspar, \eruw, and \erw consistently rank as top performers, while \gspar and \scan persistently underperform.

\subsection{High-level Metrics}
\begin{figure}[h!]
\begin{subfigure}[b]{0.5\textwidth}
    \centering
    \includegraphics[width=0.7\linewidth, trim={0 0.5cm 0 0}, clip]{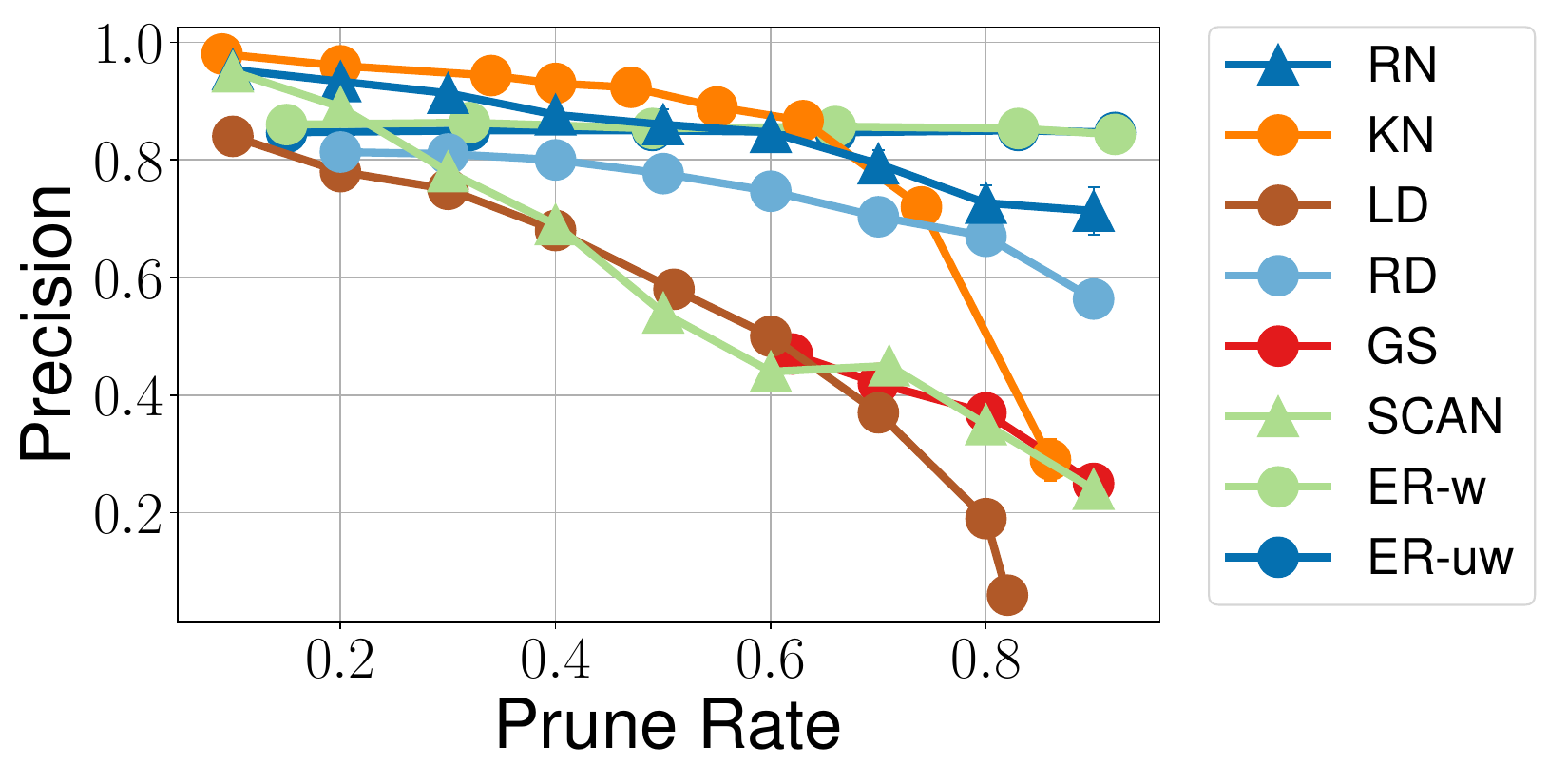}
    \caption{PageRank Centrality on \google}
    \label{fig:google_pagerank}
\end{subfigure}

\begin{subfigure}[b]{0.5\textwidth}
    \centering
    \includegraphics[width=0.7\linewidth, trim={0 0.5cm 0 0}, clip]{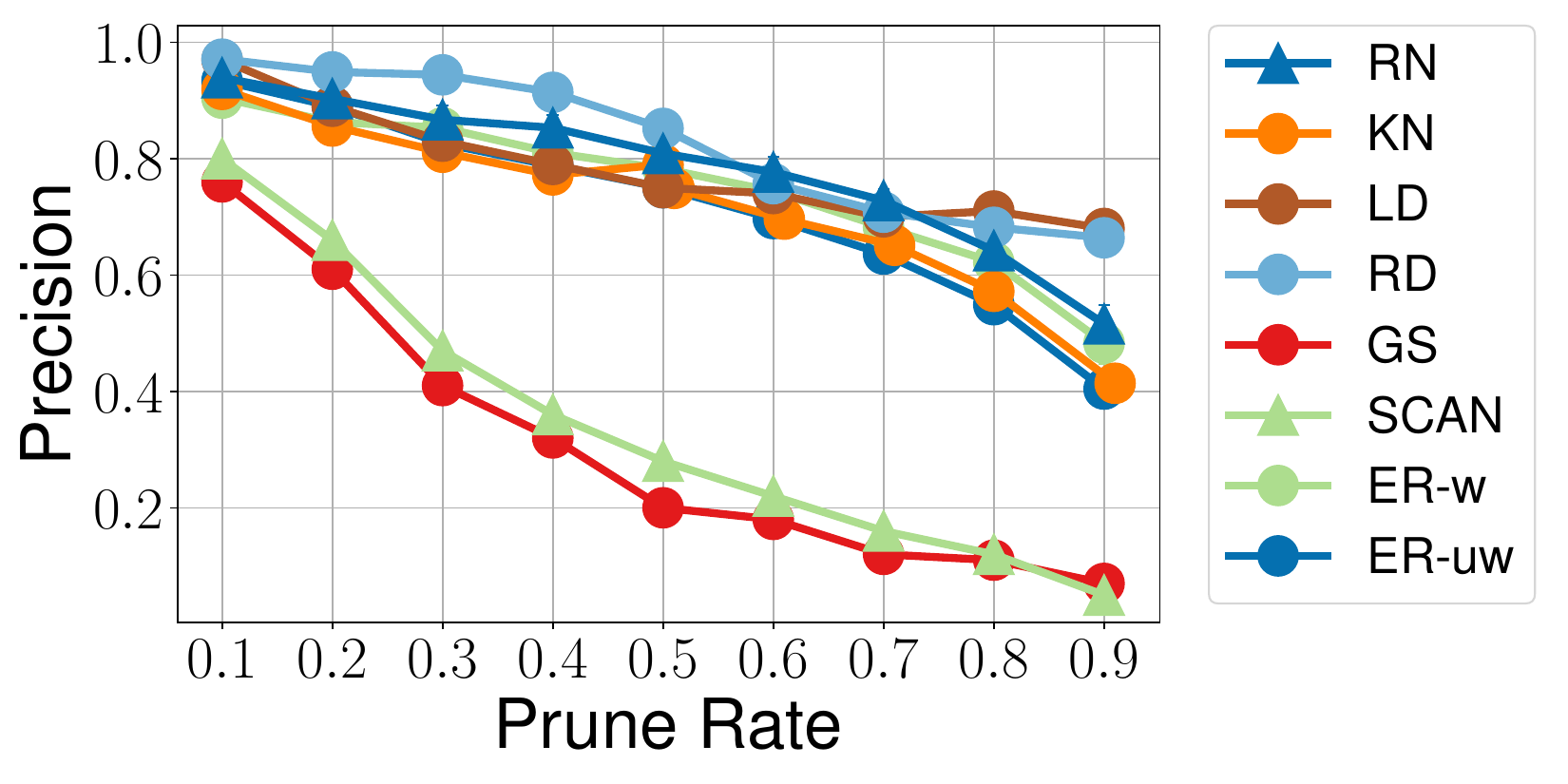}
    \caption{PageRank Centrality on \facebook}
    \label{fig:facebook_pagerank}
\end{subfigure}
    \caption{PageRank centrality. Higher precision is better. (a) PageRank centrality on \google. \kneighbor and \random perform the best at a low prune rate, \erw and \eruw perform the best at a high prune rate. \localdegree, \gspar, and \scan do not perform well. (b) PageRank centrality on \facebook. \rankdegree has the best performance. \gspar and \scan underperform.}
    \label{fig:pagerank}
\end{figure}

\textbf{PageRank.} Figures~\ref{fig:google_pagerank} and ~\ref{fig:facebook_pagerank} present the top-100 precision of PageRank on \google and \facebook, respectively. Note that \google is a directed graph and \er only supports undirected graphs. Thus, we symmetrize the graph before performing \er. Sparsifiers that work on directed graphs are applied directly.

As illustrated in Figure~\ref{fig:google_pagerank}, \eruw and \erw demonstrate high precision and consistency at various prune rates. On all web networks (\notredame, \berkstan, \google, \stanford) the performance of \er remains similar in that precision is almost constant at different prune rates. However, \er does not always achieve the best performance at low prune rates. For some graphs, the precision of \er remains constant but at a lower level. This can be due to the symmetrizing process altering the original graph's information. The more symmetrical the original graph is, the less influence will be introduced. \kneighbor also shows good performance at low prune rates. In contrast, \gspar, \scan, and \localdegree fail to preserve PageRank effectively.

Figure~\ref{fig:facebook_pagerank} reveals \er sparsifier's performance on unweighted graphs, using ego-Facebook as an example. \rankdegree, \localdegree, \random, \kneighbor, \eruw, and \erw all exhibit similar performance in preserving PageRank. \gspar and \scan continue to underperform. In comparison to directed graphs, \er no longer exhibits almost constant precision at varying prune rates on undirected graphs. \rankdegree and \kneighbor consistently perform well on both directed and undirected graphs. \localdegree displays a significant discrepancy in performance between directed and undirected graphs, excelling in undirected graphs but consistently underperforming in directed ones. \gspar and \scan show poor performance consistently.

\begin{figure}[h!]
    \centering
    \includegraphics[width=0.7\linewidth, trim={0 0.5cm 0 0}, clip]{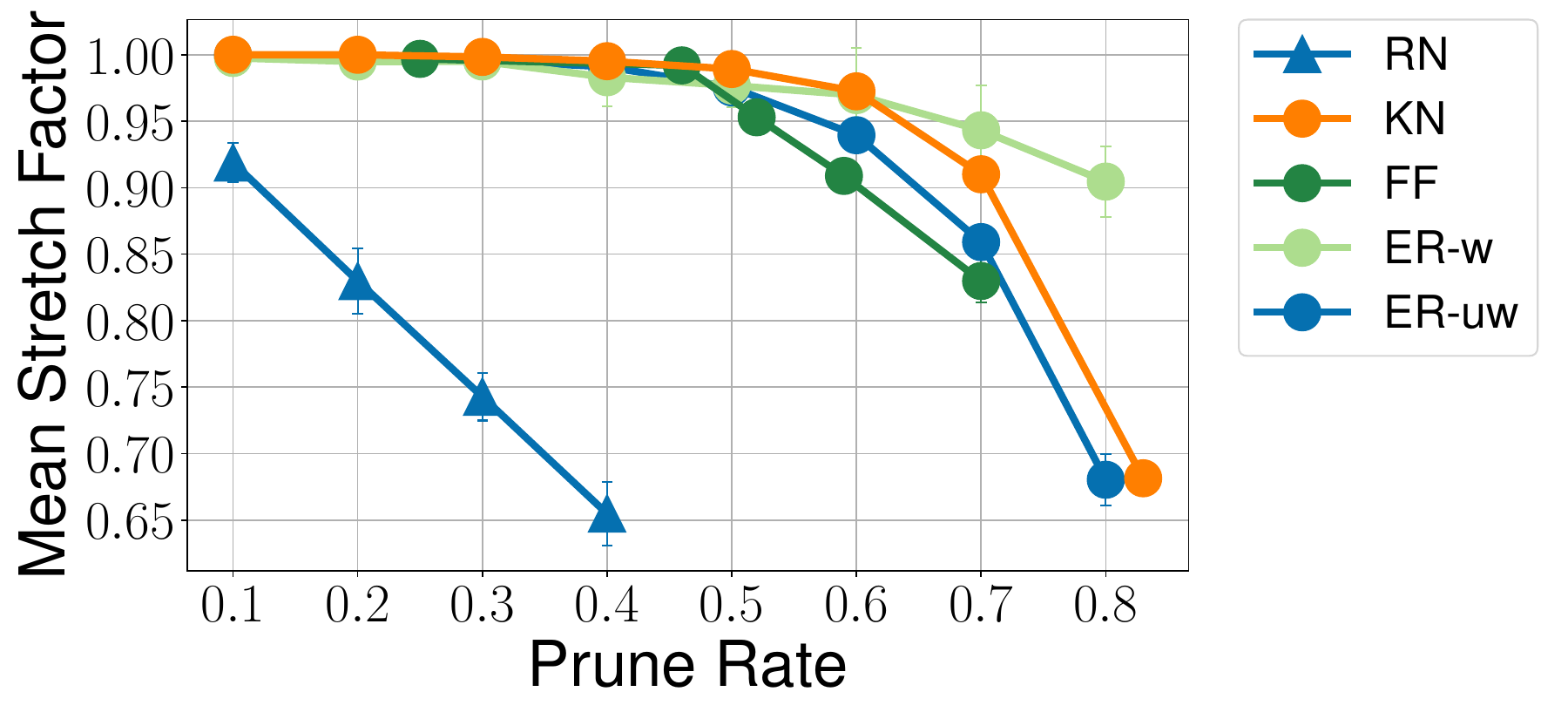}
    \caption{Adjusted Mean Stretch Factor for min-cut/max-flow with the constraint of acceptable unreachable ratio on \hepph. Closer to 1 is better. \erw has the best performance.}
    \label{fig:maxflow_stretch_factor}
\end{figure}

\begin{figure*}[h!]
\begin{subfigure}[b]{0.48\textwidth}
    \centering
    \includegraphics[width=0.7\linewidth, trim={0 0.5cm 0 0}, clip]{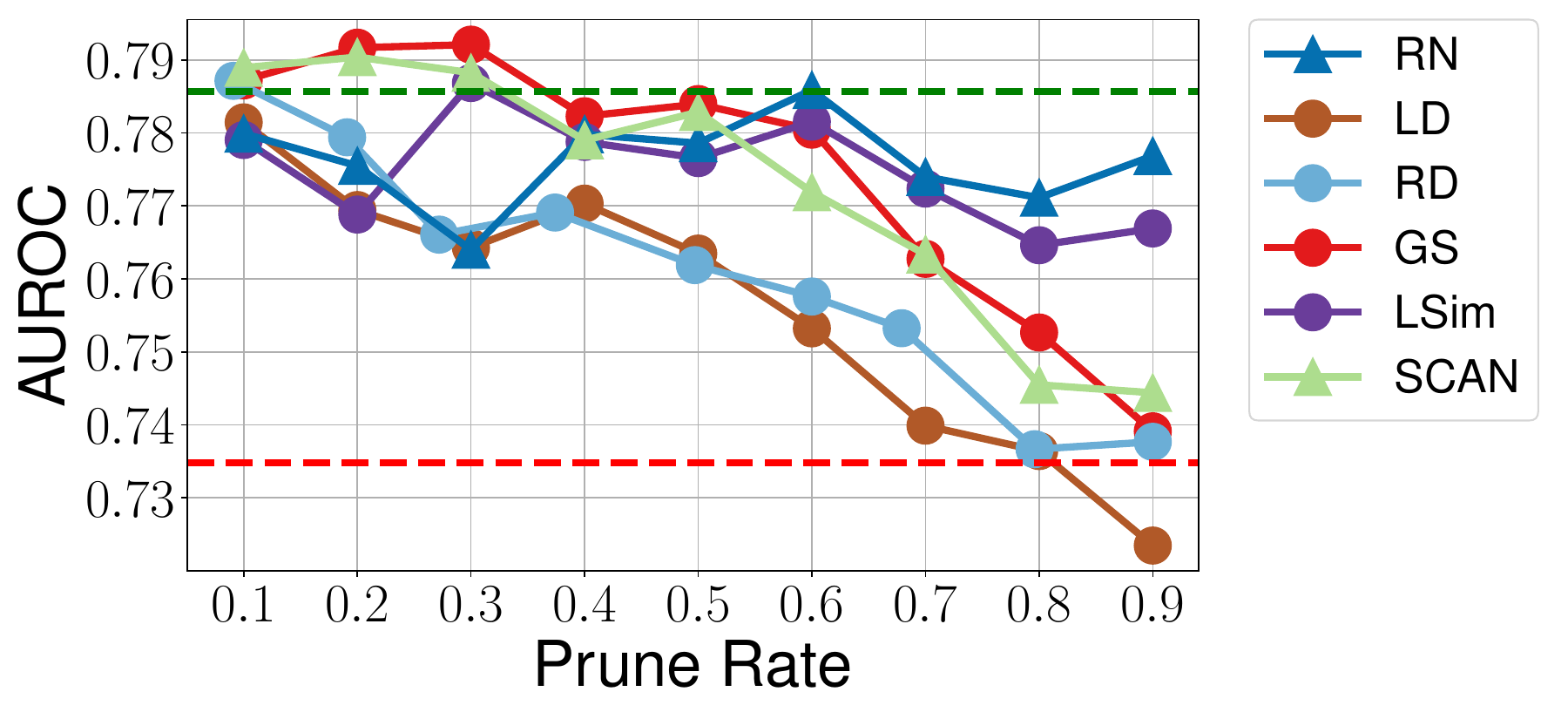}
    \caption{GraphSAGE comparison of different sparsifiers on \proteins}
    \label{fig:sage}
\end{subfigure}
\begin{subfigure}[b]{0.48\textwidth}
    \centering
    \includegraphics[width=0.7\linewidth, trim={0 0.5cm 0 0}, clip]{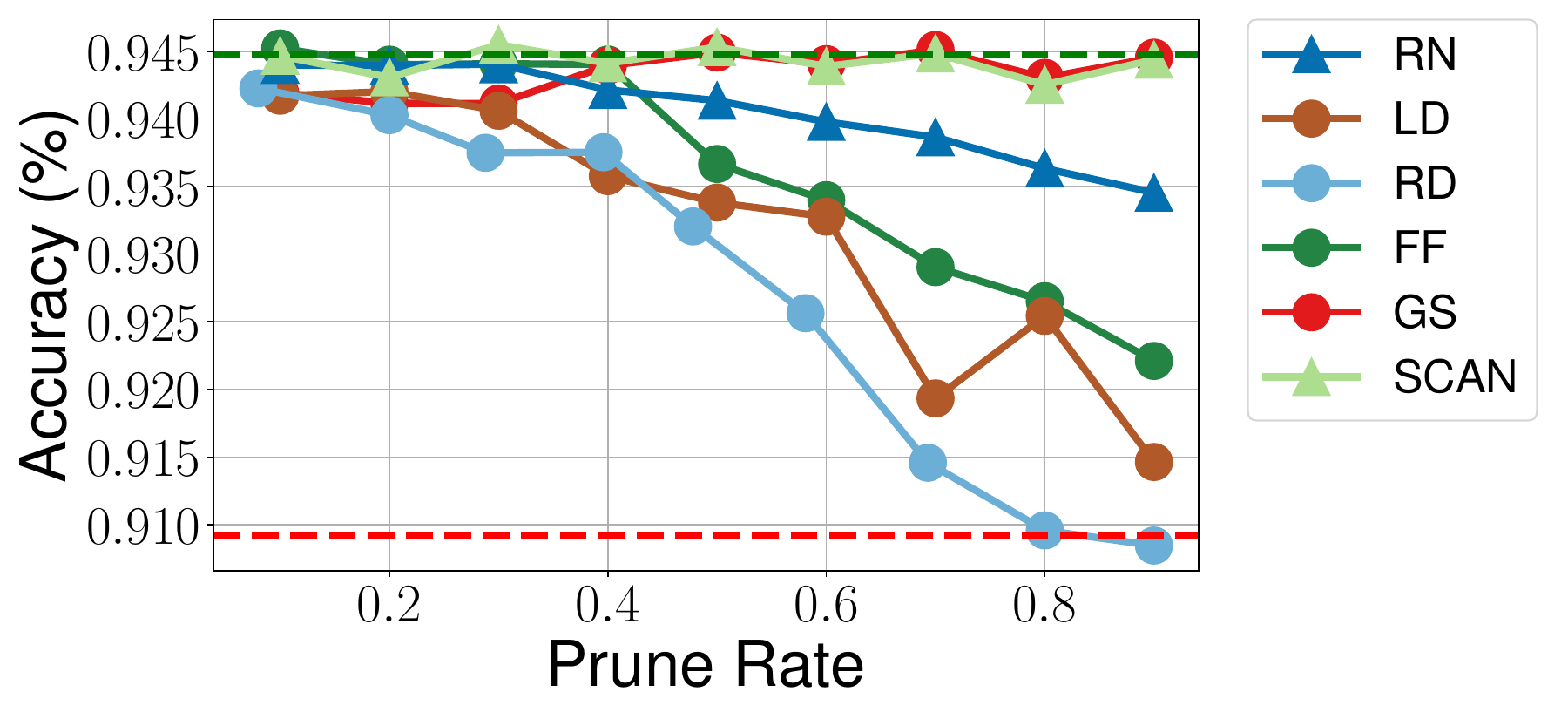}
    \caption{ClusterGCN comparison of different sparsifiers on \reddit}
    \label{fig:clustergcn}
\end{subfigure}
    \caption{GNN comparison of different sparsifiers. Higher AUROC and accuracy are better. The green line represents the inference results on the model trained by the full graph. The red line represents the inference results on the model trained with no graph (MLP only). (a) is evaluated with the GraphSAGE on \proteins. (b) is evaluated with the ClusterGCN on \reddit.}
    \label{fig:gnn}
\end{figure*}

\textbf{Min-cut/Max-flow.} 
Figure~\ref{fig:maxflow_stretch_factor} presents the mean stretch factor on \hepph, with the constraint that the unreachable ratio remains <20\% higher than in the original graph. A mean stretch factor closer to 1 means better performance. \erw shows the best performance. This can be attributed to \er being a spectral sparsifier, which preserves the spectral properties of graphs~\cite{spielman2011}. Min-cut/max-flow methods are also closely related to the graph spectrum. Flow-based graph partitioning~\cite{andrew2001} employs the Fiedler vector~\cite{Fiedler1973} (eigenvector corresponding to the second smallest eigenvalue of the graph Laplacian). One can intuitively think of \er as retaining high-resistance (low-redundant) edges in the graph, which are typically found in the critical (narrowest) section of the max-flow problem. \erw significantly outperforms its unweighted counterpart \eruw, as \erw effectively compensates for the weights of other edges when removing edges. \kneighbor and \forestfire show good empirical performance as well. In contrast, \gspar and \scan underperform, and other sparsifiers exhibit similarly mediocre results. The outcomes for min-cut/max-flow are consistent across graphs, with \erw as the top performer, followed by \kneighbor and \forestfire.

\textbf{GNN.} Figures~\ref{fig:sage} and \ref{fig:clustergcn} show the performance of sparsifiers on two distinct GNN models. GNN performance is measured using AUROC and/or vertex classification accuracy. The green dashed line represents performance on the full graph, while the red dashed line represents performance on the empty graph (a graph with no edges). We include the empty graph to demonstrate the performance of GNN models based solely on vertex features without any graph structural information. 
On the GraphSAGE model, \random and \localsimilarity perform the best; \gspar and \scan show good performance at low prune rates but deteriorate rapidly at higher rates. However, on the CluterGCN model, \gspar and \scan perform well at all prune rates. \localdegree and \rankdegree consistently underperform compared to other sparsifiers on both models. Due to the complexity of GNN algorithms, it is challenging to draw straightforward conclusions. Overall, the performance of sparsifiers on GNNs differs from model to model, which may be due to the inherent characteristics of GNN workloads.

\subsection{Sparsification Time}
\begin{figure}[h!]
    \centering
    \includegraphics[width=0.7\linewidth]{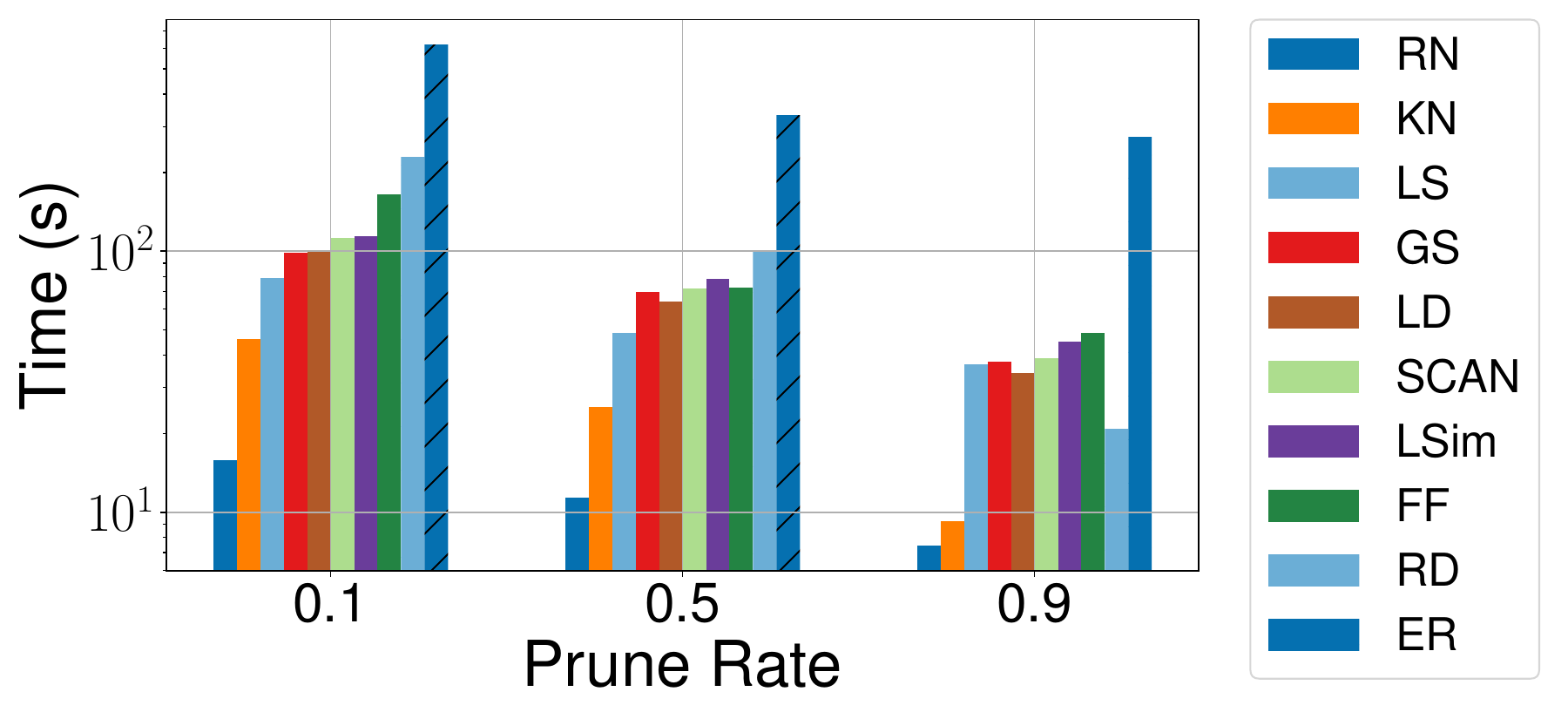}
    \caption{Sparsification time comparison on \proteins}
    \label{fig:time_bar}
\end{figure}

Figure~\ref{fig:time_bar} shows the sparsification time of different sparsifiers at different prune levels. For all sparsifiers, sparsification time decreases as the prune rate increases, this is expected because the higher the prune rate, the fewer edges need to be picked. 
Across sparsifiers, the sparsification time is also different. \random and \kneighbor are the sparsifiers with the lowest overhead due to their low algorithmic complexity.
\lspar, \gspar, \localdegree, \scan, \localsimilarity, \forestfire, and \rankdegree show similar latency.
\er is the most complex algorithm. In the figure, the time for \er is only for sampling. We do not include the computation time of the effective resistance because it is a one-time cost. The computation of effective resistance takes 990 seconds for \proteins. and the execution time of \er is approximately an order of magnitude higher than that of other sparsifiers.
However, depending on the application, a high-cost sparsifier like \er can still be useful if it preserves the desired graph properties, and the sparsification overhead is less than the time that can be saved in performing the downstream task on the sparsified graph.

\subsection{Summary of Results and Insights}
Overall, The performance of all sparsifiers degrades as the prune rate increases. Usually, we observe that the relative performance of sparsifiers is consistent across prune rates, meaning superior sparsifiers at low prune rates will remain superior at high prune rates, and the performance gap between the superiors and inferiors will be larger. 
On some occasions, the performance of a sparsifier have an elbow point, beyond which the performance drops sharply. This is due to some sparsifiers unable to maintain certain property beyond a the elbow prune rate. For example, in figure~\ref{fig:eigenvector}, the performance of \localdegree dropped abruptly when increasing the prune rate from 0.8 to 0.9, because the number of edge is so low that it cannot maintain the graph connectivity anymore.

To make sparsification effective, the selection of the sparsification algorithm should preserve the graph property/properties on which the downstream application is based. We summarize the what each sparsifier preserve below.

\begin{itemize}
    \item \textbf{\random}: preserves relative (distribution-based or ranking-based) properties, for example, degree distribution, and top centrality rankings. It struggles to preserve absolute (valued-based) properties, for example, number of communities, clustering coefficient, and min-cut/max-flow. 
    \item \textbf{\kneighbor}, \textbf{\spanning}, and \textbf{\tspanners}: preserves graph connectivity; keeps pair unreachable ratio and vertex isolated ratio low.
    \item \textbf{\rankdegree} and \textbf{\localdegree}: preserves graph connectivity and edges to high-degree vertices (hub vertices). Perform well on distance metrics (APSP, eccentricity, diameter) and centrality metrics.
    \item \textbf{\forestfire}: simulates the evolution of graphs and does not strictly stick to the original graph. Empirically it does not excel at any metrics evaluated.
    \item \textbf{\gspar} and \textbf{\scan}: Empirically perform well in preserving ClusterGCN accuracy.
    \item \textbf{\lspar} and \textbf{\localsimilarity}: preserves the edge to similar vertices, thus preserves clustering similarity.
    \item \textbf{\er}: preserves the spectral properties of the graph, specifically the quadratic form of the graph Laplacian. It perform well in preserving min-cut/max-flow results.
\end{itemize}

\section{Related Work}

ML-based sparsifiers are a group of sparsifiers that use machine learning-related techniques to sparsify graphs. SparRL~\cite{wickman2022} proposes a graph sparsification framework enabled by deep reinforcement learning. NeuralSparse~\cite{zheng2020} presents a supervised graph sparsification technique to improve performance in graph neural networks (GNN). Instead of focusing on saving execution time by performing graph sparsification, NeuralSparse aims to remove task-irrelevant edges from the graph, and thus improve the accuracy of the downstream GNNs. DropEdge~\cite{rong2019} presents a method very close to the random sparsifier, but samples a random set of edges for each training epoch in graph convolutional network (GCN), the goal is both to reduce message passing overhead and reduce over-fitting with the full graph input.

\section{Conclusion}
This study provides a comprehensive evaluation of 12 graph sparsification algorithms, analyzing their performance in preserving 16 essential graph metrics across 14 real-world graphs with diverse characteristics. Our findings revealed that no single sparsifier excels in preserving all graph properties, and it is important to select appropriate sparsification algorithms based on the downstream task. This study contributes to the broader understanding of graph sparsification algorithms, and we provided insights to guide future work in effectively integrating graph sparsification into graph algorithms to optimize computational efficiency without significantly compromising output quality. Our open-source framework offers a valuable resource for ongoing evaluations of emerging sparsification algorithms, graph metrics, and growing graph data.

\begin{acks}
This research is based upon work supported by the Office of the Director of National Intelligence (ODNI), Intelligence Advanced Research Projects Activity (IARPA), through the Advanced Graphical Intelligence Logical Computing Environment (AGILE) research program, under Army Research Office (ARO) contract number W911NF22C0085. The views and conclusions contained herein are those of the authors and should not be interpreted as necessarily representing the official policies or endorsements, either expressed or implied, of the ODNI, IARPA, ARO, or the U.S. Government.  This work was also supported by the United States-Israel BSF grant number 2020135.
\end{acks}

\bibliographystyle{ACM-Reference-Format}
\newpage
\bibliography{99_references}

\end{document}